\documentclass[journal]{IEEEtran}

\usepackage{threeparttable}
\usepackage{hyperref}
\usepackage{url}
\usepackage{amsmath,amssymb}
\usepackage{enumerate}

\newcounter{MYtempeqncnt}

\newcommand{\beq}{\begin{equation}}
\newcommand{\beql}[1]{\begin{equation}\label{#1}}
\newcommand{\eeq}{\end{equation}}

\usepackage{etoolbox}
\IEEEoverridecommandlockouts
\usepackage{cite}
\usepackage{amsmath,amssymb,amsfonts}

\usepackage{tkz-euclide,tikzscale}
\usetikzlibrary{arrows,circuits.ee.IEC}
\tikzset{ac source/.style={
  circuit symbol lines,
  circuit symbol size = width 2 height 2,
  shape = generic circle IEC,
  /pgf/generic circle IEC/before background={
    \pgfpathmoveto{\pgfpoint{-0.8pt}{0pt}}
    \pgfpathsine{\pgfpoint{0.4pt}{0.4pt}}
    \pgfpathcosine{\pgfpoint{0.4pt}{-0.4pt}}
    \pgfpathsine{\pgfpoint{0.4pt}{-0.4pt}}
    \pgfpathcosine{\pgfpoint{0.4pt}{0.4pt}}
    \pgfusepath{stroke}
  },
  transform shape
}}

\usepackage{graphicx}
\usepackage{textcomp}
\usepackage{xcolor}
\usepackage{graphicx}      
\usepackage{bm}
\usepackage{epsfig}
\usepackage{epstopdf}
\usepackage{float}
\usepackage{mathrsfs}
\usepackage{xtab}
\usepackage{color}
\usepackage{ifthen}
\usepackage{setspace}
\usepackage{algorithm}
\usepackage{algorithmic}
\usepackage{dsfont}
\usepackage{booktabs}
\usepackage{verbatimbox}
\usepackage{subfig}
\usepackage{ntheorem}

\theoremseparator{:}
\newtheorem{theorem}{Theorem}

\newtheorem{corollary}{Corollary}
\newtheorem{remark}{Remark}
\usepackage[font=sc,labelsep=period]{caption}
\usepackage{multirow}

\begin{document}

	\title{A Scenario-oriented Approach to Energy-Reserve Joint Procurement and Pricing}
	
	\author{Jiantao Shi,\IEEEmembership{~Student~Member,~IEEE,}
		Ye Guo,\IEEEmembership{~Senior~Member,~IEEE,} Lang Tong,\IEEEmembership{~Fellow,~IEEE,} Wenchuan Wu,\IEEEmembership{~Fellow,~IEEE,}
		and Hongbin Sun,\IEEEmembership{~Fellow,~IEEE}
		\thanks{Part of the work was presented at the 2021 IEEE PES General Meeting \cite{PESGM}.
		
		The work was supported in part by the National Science Foundation of China under Grant 51977115. (Corresponding author:
        Ye Guo, e-mail: guo-ye@sz.tsinghua.edu.cn)

        Jiantao Shi, Ye Guo, Wenchuan Wu and Hongbin Sun are with the Smart Grid and Renewable Energy Laboratory, Tsinghua-Berkeley Shenzhen Institute, Shenzhen, 518071 China. Wenchuan Wu and Hongbin Sun are also affiliated with the State Key Laboratory of Power Systems, Department of Electrical Engineering, Tsinghua University, Beijing, 100084 China.
        
        Lang Tong is with the School of Electrical and Computer Engineering, Cornell University, Ithaca, NY 14853, USA.}}

	
	\maketitle
	
	\begin{abstract}
    We consider some crucial problems related to the secure and reliable operation of power systems with high renewable penetrations: how much reserve should we procure, how should reserve resources distribute among different locations, and how should we price reserve and charge uncertainty sources. These issues have so far been largely addressed empirically. In this paper, we first develop a scenario-oriented energy-reserve co-optimization model, which directly connects reserve procurement with possible outages and load/renewable power fluctuations without the need for empirical reserve requirements. Accordingly, reserve can be optimally procured system-wide to handle all possible future uncertainties with the minimum expected system total cost. Based on the proposed model, marginal pricing approaches are developed for energy and reserve, respectively. Locational uniform pricing is established for energy, and the similar property is also established for the combination of reserve and re-dispatch. In addition, properties of cost recovery for generators and revenue adequacy for the system operator are also proven.
	\end{abstract}
	
	\IEEEpeerreviewmaketitle

	\section{Introduction}
	\subsection{Backgrounds}
	\IEEEPARstart {T}{he} integration of more renewable generations in power systems is an important way to achieve carbon neutrality. However, the increasing uncertainty brought by renewable generations also brings challenges to the secure and reliable operation of power systems. To handle this, the reserve is procured system-wide and deployed for generation re-dispatch when contingencies happen, or loads/renewable generations deviate from their predictions. Reserve and energy are strongly coupled both in generation and transmission capacity limits. A fair, efficient, and transparent energy-reserve co-optimization model and the associated market mechanism are of crucial importance. 
	
	\subsection{Literature Review}
	Energy and reserve markets are cleared either sequentially or jointly to deal with their coupling in generation capacities. Currently, most of the independent system operators (ISOs) in the U.S. adopt a joint clearing process\cite{review1,review2}, stylistically defined by:
	\begin{align} 
	\label{obj1}
	& \mbox{(I)}: \quad\underset{\{g,r_U,r_D\}}{\rm minimize} \quad c_{g}^T g + c_{U}^T r_U + c_{D}^T r_D\\
	&\mbox{subject to}\notag\\
	\label{traditional balance and pf}
	&(\lambda,\mu) : \mathds{1}^T g =  \mathds{1}^T d, S(g-d)\leq f,\\
	\label{reserve requirement}
	&(\gamma^U,\gamma^D):\mathds{1}^T r_U =  R^U,\mathds{1}^T r_D =  R^D,\\
	\label{physical limit}
	&g + r_U \leq \overline{G}, \underline{G}+ r_D \leq g,0 \leq r_U \leq \overline{r_U}, 0 \leq r_D \leq \overline{r_D}.
	\end{align}
	The objective function (1) aims to minimize the total bid-in cost of energy $g$, upward reserve $r_U$, and downward reserve $r_D$. Constraints (2)-(4) represent energy balancing and transmission capacity constraints, reserve requirement constraints, and generator capacity and ramping-rate limits, respectively. Reserve clearing prices will be set as the Lagrange multipliers $(\gamma^U,\gamma^D)$, representing the increase of total bid-in cost when there is one additional unit of upward/downward reserve requirement.
	
	Several important issues arise from model (I). First, the reserve requirements $R^U$ and $R^D$ are empirical and subjective, sometimes specified as the capacity of the biggest online generator as in PJM\cite{PJM} or a certain proportion of system loads as in CAISO\cite{CAISO}. Their values, however, can significantly affect the market clearing results and prices of both the energy and reserve products. Second, the deliverability of reserve in non-base scenarios with contingencies and load/renewable power fluctuations is not considered in (I). A common solution is to partition the entire system into different zones, each with zonal reserve requirements and prices\cite{ISO-NE}, which is again ad hoc. Third, the objective function (\ref{obj1}) uses the base-case bid-in cost without taking into account possible re-adjustment costs. Fourth, it is not clear how the cost of procuring reserve should be afforded among all consumers. Some ISOs let all load serving entities share this cost proportionally\cite{review1,review2}.
	
	In light of these problems, many studies aim to improve the traditional model (I). For reserve requirement selection, parametric and non-parametric probabilistic forecasting techniques are incorporated in (I) by characterizing the underlying probability distribution of future situations \cite{para1,para2,nonpara1,nonpara2}. Scenario-based approaches have also been proposed in\cite{scen1,scen2}. In addition, on the deliverability of reserve, a statistical clustering method is proposed in \cite{area3} that partitions the network into reserve zones. In \cite{diaqual}, a generalized reserve disqualification approach is developed. 
	
	Stochastic co-optimization models have also been proposed. In \cite{UMP}, a robust stochastic optimization model is adopted for the co-optimization of energy and possible re-dispatch. Moreover, the uncertainty marginal price is defined as the marginal cost of immunizing the next unit increment of the worst point in the uncertainty set. Such marginal cost is used to price both reserve resources and uncertainty sources. In \cite{Yury2020}, a chance-constrained stochastic optimization model is adopted, where statistical moments of uncertainties are considered to generate chance constraints, and an algorithm is proposed to transform the original problem to a convex formulation. In \cite{Morales2012,Morales2014,Kazempour2018,Wong2007,readjust1,readjust2,readjust3,reserve_LMP_1}, a scenario-oriented stochastic optimization model structure is adopted. Some non-base scenarios with occurrence probabilities are modelled to represent possible loads/renewable power fluctuations and contingencies. The energy balancing and transmission capacity constraints in all non-base scenarios are considered to analyze the re-adjustment procedures. Among those scenario-based solutions, some adopt the energy-only model structure \cite{Morales2012,Morales2014,Kazempour2018}; others consider the energy-reserve co-optimization structure \cite{Wong2007,readjust1,readjust2,readjust3,reserve_LMP_1}. In \cite{readjust1}, the nodal marginal prices of energy and security are derived to settle energy and all tiers of regulating reserve, respectively.
	
	On the design of a pricing approach, the revenue adequacy of the operator and the cost recovery of market participants are necessary. In \cite{Morales2012} and \cite{Wong2007}, both properties are established in expectation. In \cite{Morales2014}, cost recovery is established in expectation, and revenue adequacy is established for every scenario. In \cite{Kazempour2018}, an equilibrium model is adopted to achieve both revenue adequacy and cost recovery for each scenario with increased system costs and the loss of social welfare.

	\subsection{Contributions, Organizations and Nomenclature}
	The main contributions of this paper are listed as follows:
	
	(1) A scenario-oriented energy-reserve co-optimization model is proposed. Reserve procured from a generator is modelled as the maximum range of its power re-dispatch in all scenarios. The proposed model no longer relies on empirical parameters of (zonal) reserve requirements. The deliverability of all reserve resources under all scenarios is ensured by incorporating the network constraints in non-base scenarios into the co-optimization model.
 	
	(2) Marginal prices of energy and reserve are derived as by-products of the co-optimization model. The associated settlement process is also presented. We show that energy prices are locational uniform, and a proportional uniform pricing property can be established for re-dispatch.
	
	(3) Cost recovery for generators is established for every scenario, and revenue adequacy for the system operator is established in expectation. We show that revenue from load payments, credits to generators including energy, reserve, and re-dispatches, and congestion rent will reach their balance for both the base case and each non-base scenario. 
	
	
	
	
	The rest of this paper is organized as follows. The proposed model is formulated in Section II. The pricing approach and the settlement process are presented in Section III. Some properties are established in Section IV. Some of the assumptions are further discussed in Section V. Case studies are presented in Section VI. Section VII concludes this paper. A list of major designated symbols are listed in Table I.
	
	{\small
\begin{table}[h]
\begin{center}
\caption{\small List of Major Symbols Used}
\begin{tabular}{|ll|}
\hline
$c_{g},c_{U},c_{D}$: & bid-in prices of energy, upward reserve\\
 &and downward reserve.\\
$\overline{c},\underline{c},c_L$: & upward/downward generation re-dispatch \\
& prices and load shedding prices.\\
$d,\pi_k$: & basic load/load fluctuation in scenario $k$.\\
$\delta d_k$: & load shedding in scenario $k$.\\
$\delta g^U_k,\delta g^D_k$: & generation upward and downward \\  & re-dispatches in scenario $k$.\\
$\epsilon_k$: & occurrence probability of scenario $k$.\\
$\eta^g,\eta^d$: & energy marginal prices of generators and loads.\\
$\eta^U,\eta^D$: & upward and downward reserve marginal prices.\\
$f_0,f_k$: & transmission capacity limit in the base case \\
& and in scenario $k$.\\
$F(\cdot)$: & objective function.\\
$g,r_U,r_D$: & generations, upward and downward reserve.\\
$\omega_0,\omega_k$: & base-case/non-base component of energy  \\
&prices.\\
$S,S_k$: & shift factor matrices in the base case\\
& and in scenario $k$\\\hline
\end{tabular}
\end{center}
\end{table}
}

	\section{Model Formulation}

\begin{figure}[H]
		\centering
	\vspace{-0.1cm}
		\includegraphics[width=3.3in]{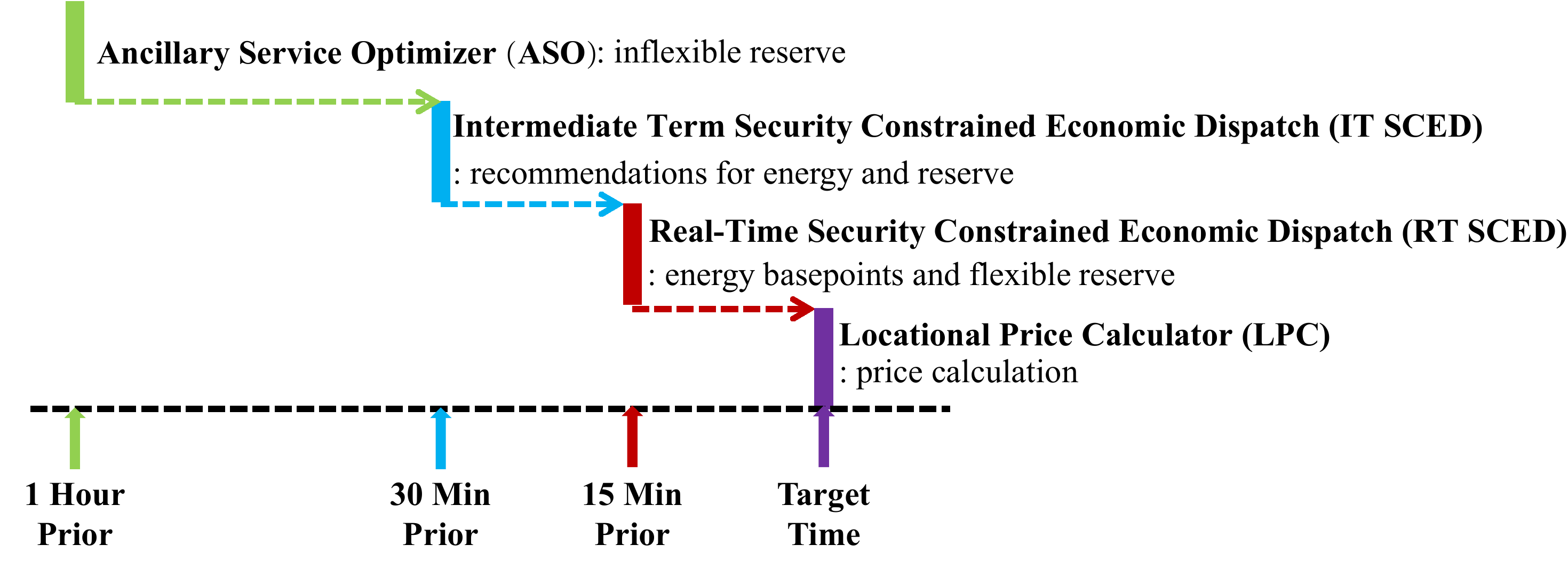}
    \setlength{\abovecaptionskip}{-3.0pt}
    \captionsetup{font=footnotesize,singlelinecheck=false}
		\caption{PJM real-time schedules of operation reserve including ASO, IT SCED, RT SCED and LPC\cite{PJM}}
		\label{PJM RT SCED}
    \vspace{-0.25cm}
	\end{figure}

	We consider a scenario-oriented co-optimization model. In the current market design, with PJM as an example, the reserve market consists of several look-ahead stages to procure reserve resources with different flexibility levels as shown in Fig. \ref{PJM RT SCED}. We abstract the real-time operations in Fig. \ref{PJM RT SCED} into a look-ahead energy-reserve co-optimization model with the following assumptions: 
    
	i) A shift factor-based lossless DCOPF model with linear cost functions for energy and reserve is adopted, which is consistent with the current electricity market design.

	ii) A single-period problem is considered for simplicity.
		
	iii) Renewable generations are modelled as negative loads.	
	
	iv) Non-base scenarios may have line outages, load or renewable power fluctuations, etc. Generator outages, however, are not considered.
		
	v) The objective of the proposed model is to guard against a set of mutually exclusive non-base scenarios, and the occurrence probability of each scenario is assumed to be given. 
	
	We will have further discussions on assumptions (ii)-(iv) in Section V. 
	
	Based on assumptions (i)-(v), the proposed co-optimization model is given by:
	\begin{align}
	&\!F(g,r_U,r_D,\delta g^U_k,\delta g^D_k,\delta d_k)=\notag\\
	\label{obj2}
	&\!c_{g}^T\!g\!+\!c_{U}^T\!r_U\!+\!c_{D}^T r_D\!+\!\sum_{k\in \mathcal{K}} \!\epsilon_k (\overline{c}^T\!\delta g^U_k \!-\!\underline{c}^T\!\delta g^D_k \!+\!c^T_{L}\!\delta d_k),\\
	&\mbox{(II)}: \quad \underset{\{g,r_U,r_D,\delta g^U_k,\delta g^D_k,\delta d_k\}}{\rm minimize} F(\cdot), \notag\\
	&\mbox{subject to}\notag\\	
	\label{base balance and pf}
	&(\lambda,\mu):\mathds{1}^T g =  \mathds{1}^T d, S_G \cdot g-S_D \cdot d\leq f,\\
	\label{base physical limit1}
	&(\underline{\upsilon},\overline{\upsilon}): \underline{G}+r_D \leq g,g + r_U \leq \overline{G},\\
	\label{base physical limit2}
	&(\underline{\rho^U},\overline{\rho^U},\underline{\rho^D},\overline{\rho^D}):0 \leq r_U \leq \overline{r_U}, 0 \leq r_D \leq \overline{r_D},\\
	&\mbox{for all $k\in \mathcal{K}$:} 
	\notag\\
	\label{cntg balance}
	&\lambda_k:\mathds{1}^T(g+\delta g^U_k-\delta g^D_k)=\mathds{1}^T (d+\pi_k -\delta d_k),\\
	\label{cntg pf}
	&\mu_k:S_{G,k} (g+\delta g^U_k-\delta g^D_k)-S_{D,k} (d+\pi_k -\delta d_k) \leq f_k,\\
	\label{dg1fanwei}
	&(\underline{\alpha_k},\overline{\alpha_k}): 0 \leq \delta g^U_k \leq r_U,\\
	\label{dg2fanwei}	
	&(\underline{\beta_k},\overline{\beta_k}): 0 \leq \delta g^D_k \leq r_D,\\
	\label{dk1fanwei}
	&(\underline{\tau_k},\overline{\tau_k}): 0 \leq \delta d_k \leq d+\pi_k,
	\end{align}
    where the objective function (\ref{obj2}) is the expected system total cost, including the base-case bid-in cost as in (1) and the expectation of re-adjustment costs in all scenarios\footnote{It should be noted that coefficients $\overline{c}$ and $\underline{c}$ in re-adjustment costs respectively represent the true marginal cost of upward re-dispatch and the true marginal cost reduce of generation downward re-dispatch. In some papers e.g. \cite{Morales2012}, $\overline{c}$ and $\underline{c}$ are set as the energy bid-in prices $c_g$. In this paper, in case some generators may incur extra costs for fast ramping, we set $\overline{c}$ and $\underline{c}$ as independent coefficients from $c_g$. Note that no matter how we set these two coefficients, all the qualitative analyses in this paper will always hold.}. If one generator's downward reserve is deployed, its output will decrease, so will its generation cost. Therefore, there is a negative sign before the term $(\underline{c}^T\delta g^D_k)$ in (\ref{obj2}). Constraints (\ref{base balance and pf})-(\ref{base physical limit2}) are base-case constraints in the same form as (\ref{traditional balance and pf})-(\ref{physical limit}) in model (I), except for the shift factor matrices used in network constraints. The two matrices, $S_{G}$ and $S_{D}$ are shift factor matrices associated with generators and loads. Constraints (\ref{cntg balance})-(\ref{cntg pf}) are the energy balancing constraints and transmission capacity limits in all non-base scenarios. Note that transmission capacities in non-base scenarios $f_k$ may not be the same as the base-case $f$. The impact of line outages on network topology is reflected in $S_{G,k}$ and $S_{D,k}$ compared with the base-case $S_{G}$ and $S_{D}$. Constraints (\ref{dg1fanwei})-(\ref{dg2fanwei}) indicate that the procured reserve will be modelled as the maximum range of generation re-dispatches in all scenarios. In constraints (\ref{cntg balance})-(\ref{cntg pf}) and (\ref{dk1fanwei}), $\pi_k$ is the vector of load fluctuation parameter in non-base scenario $k$ forecast by the system operator. Therefore, constraint (\ref{dk1fanwei}) represents that load shedding in each scenario must be non-negative and cannot exceed forecast load power $d+\pi_k$ in scenario $k$.\footnote{We assume that $\delta d_k$ and $d+\pi_k$ are non-negative. If they are negative for some resources in scenario $k$, it means that these resources are uncontrollable renewable generations and $\delta d_k$ represents renewable curtailment. To consider this, we only need to add one more constraint $d\!+\!\pi_k\! \leq \!\delta d_k \!\leq \! 0$ and one associated dual variable, and all the qualitative analyses will still hold.}
    
    
    In some market implementations, the one-step ramping constraint from the previous dispatch set point will be considered as $g^{SE}-\overline{r_D} \leq g \leq g^{SE}+\overline{r_U}$, where $g^{SE}$ is actual generator outputs at the last interval from state estimation, and $\overline{r_U},\overline{r_D}$ are the maximum upward and downward ramping rates of generators\cite{PJMSCED}. Note that this constraint can be easily incorporated into the generation capacity limit (\ref{base physical limit1}).

    In addition, the proposed co-optimization model (\uppercase\expandafter{\romannumeral2}) is a standard linear programming. In this paper, we do not go into its detailed solution method, except mentioning that some distributed optimization techniques can be employed to solve it efficiently, see \cite{riskSCED} and \cite{On_robust}. 
	
	Moreover, from the proposed model (II) we can derive marginal prices of energy and reserve, and design the associated market settlement process, as in the next section.

	\section{Pricing and Settlement}
	\subsection{Energy and Reserve Prices}
	The proposed pricing approach for energy and reserve is based on their marginal contributions to the expected system total cost in (\ref{obj2}). Namely, consider any generator $j$, we first fix $g(j),r_U(j),r_D(j)$ at their optimal values $g^*(j),r_U^*(j),r_D^*(j)$ and consider them as parameters instead of decision variables. Such a modified model is referred to as model (III) 
	\begin{equation}
	\setlength\abovedisplayskip{1pt}
	\setlength\belowdisplayskip{1pt}
	\underset{x_{-j} \in \mathcal{X}_{-j}}{\rm minimize} \quad F_{-j}(x_{-j}),\notag
	\end{equation}
	where $x_{-j}$ represent all the decision variables of model (II) except for $g(j),r_U(j)$ and $r_D(j)$, $F_{-j}(\cdot)$ is the overall cost excluding the bid-in cost of generator $j$, and $\mathcal{X}_{-j}$ are the constraints (\ref{base balance and pf})-(\ref{dk1fanwei}) excluding the $j^{th}$ row of all constraints in (\ref{base physical limit1})-(\ref{base physical limit2}), which are the internal constraints of generator $j$. Subsequently, we evaluate the sensitivity of the optimal objective function of model (III) $F_{-j}(x_{-j}^*)$, which represents the sensitivity of the expected cost of all other market participants except for generator $j$, with respect to parameters $g(j),r_U(j)$ and $r_D(j)$. According to the envelope theorem, the marginal energy price of generator $j$ is
	\begin{align}
	\label{RGMP2}
	\eta^g(j)& \!=\!-\frac{\partial F_{-j}(x_{-j}^*)}{\partial g(j)}\!\notag\\
	&=\!\lambda^*\!-\!S_G(\cdot,m_j)^T\!\mu^*\!+\!\sum_{k\in \mathcal{K}}\!(\lambda_k^*\!-\!S_{G,k}(\cdot,m_j)^T\!\mu_k^*)\notag\\
	&=\omega_0(m_j)+\sum_{k\in \mathcal{K}}\omega_k(m_j),
	\end{align}
	where $m_j$ is the index of the bus where generator $j$ is located. The term $(\lambda-S_G(\cdot,m_j)^T\mu)$, denoted by $\omega_0(m_j)$, corresponds to the base-case contribution to the energy prices of generators. The term $(\lambda_k-S_{G,k}(\cdot,m_j)^T\mu_k)$, denoted by $\omega_k(m_j)$, corresponds to the contribution to the generator energy prices of scenario $k$. They both follow the same form as the standard LMP, as the sum of an energy component and a congestion component. 
	
	Similarly, the energy marginal price of load $l$ is
	\begin{align}
	\label{L/d}
	\eta^d(l)&=\!\frac{F_{-j}(x_{-j}^*)}{\partial d(l)}\notag\\
	&=\!\lambda^*\!-\!S_D(\cdot,m_l)^T\!\mu^*\!+\!\sum_{k\in \mathcal{K}}(\lambda_k^*\!-\!S_{D,k}(\cdot,m_l)^T\!\mu_k^*\!-\!\overline{\tau}_k^*\!(l))\notag\\
	&=\omega_0(m_l)+\sum_{k\in \mathcal{K}}\omega_k(m_l)-\sum_{k\in \mathcal{K}} \overline{\tau}_k^*(l),
	\end{align}
	which is consistent with the energy marginal prices of generators in (\ref{RGMP2}) except for the last term $(-\!\sum\!\overline{\tau}_k^*(l))$. Here $\sum\!\overline{\tau}_k^*(l)$ are the multipliers associated with the upper bound of load shedding in (\ref{dk1fanwei}). $\overline{\tau}_k^*(l)$ is non-zero only if load $l$ will be totally shed in scenario $k$. If a load is completely shed for a particular customer whereas other loads are not, then in fact these consumers have different reliability priorities. In other words, electricity is no longer a homogeneous good for all customers. Nevertheless, we consider such cases rarely happen and introduce the following assumption in some of our analysis:
	
	vi): $\sum\!\overline{\tau}_k^*(l)$ are assumed to be zero in (\ref{L/d}).
	
	Although the energy prices of generators and loads are defined separately in (\ref{RGMP2}) and (\ref{L/d}) at the resource level, with the new assumption (vi), the property of locational uniform pricing can be established for energy as follows:

	\begin{theorem}[Locational Uniform Pricing for Energy]
	    Consider any two generators $i,j$ and any load $l$ at the same bus, i.e., $m_i\!=\!\!m_j\!=\!m_l$. Under assumptions (i)-(vi), they have the same energy price, i.e., $\eta^g(i)\!=\!\eta^g(j)\!=\!\eta^d(l)$. 
	\end{theorem}

	Furthermore, for the upward reserve marginal price, according to the envelop theorem, there is
	\begin{equation}
	\label{RUMP}
	\eta^U(j)=-\frac{\partial F_{-j}(x_{-j}^*)}{\partial r_U(j)}=\sum_{k\in \mathcal{K}} \overline{\alpha_k}^*(j).
	\end{equation} 
	From (\ref{RUMP}) we can see that for each generator $j$, if its upward generation re-dispatch reaches its procured reserve in scenario $k$, i.e., $\delta g^U_k(j)\!=\!r_U(j)$, then the corresponding multiplier $\overline{\alpha_k}^*(j)$ may be positive and contribute to its upward reserve marginal price $\eta^U(j)$. The upward reserve marginal price in (\ref{RUMP}) is discriminatory because reserve capacities procured from different generators at the same bus may not be homogeneous good due to their possible different re-dispatch prices.
	
	Similarly, the downward reserve marginal price is
	\begin{equation}
	\label{RDMP}
	\eta^{D}(j)=\sum_{k\in \mathcal{K}}\overline{\beta_k}^*(j),
	\end{equation}
	which is also discriminatory.

	Based on the proposed pricing approach, the market settlement process will be presented in the next subsection.
	
	\subsection{Market Settlement Process}
	The proposed settlement process includes two stages: in the ex-ante stage, without knowing which scenario will happen, we solve the co-optimization problem (II) to guard against all possible non-base scenarios; and in the ex-post stage, with the realization of one specific scenario, re-adjustment strategies will be deployed according to the results from model (II), and the ex-post settlement depends on the realized scenario.
	\subsubsection{ex-ante stage}
    Settlement in the ex-ante stage includes the following credits and payments: 
	\begin{itemize}
		\item Base-case contribution to generator $j$'s energy credit: 
		\begin{equation}
		\label{base-case prices to generator energy credit}
		\Gamma^g_0(j)=\omega_0(m_j) g(j);
		\end{equation}
		
		\item Non-base contributions to generator $j$'s energy credit:
		\begin{equation}
		\label{non-base scenario prices to generator energy credit}
		\sum_{k\in \mathcal{K}}\Gamma^g_k(j)=\sum_{k\in \mathcal{K}}\omega_k(m_j) g(j);
		\end{equation}
		
		\item Base-case contribution to load $l$'s energy payment:
		\begin{equation}
		\label{base-case prices to load payment}
		\Gamma^d_0(l) = \omega_0(m_l) d(l);
		\end{equation}
		
		\item Non-base contributions to load $l$'s energy payment:
		\begin{equation}
		\label{non-base scenario prices to load payment in all scenarios}
		\sum_{k\in \mathcal{K}}\Gamma^d_k(l)= \sum_{k\in \mathcal{K}}\omega_k(m_l) d(l);
		\end{equation}
		
	\item Load $l$'s fluctuation payment:
		\begin{align}
		\label{fluctuation payment}
			\sum_{k\in \mathcal{K}}\Gamma^{\pi}_k(l)
			=\sum_{k\in \mathcal{K}} \frac{\partial F_{-j}(x_{-j}^*)}{\partial \pi_k(l)} \pi_k(l)=\sum_{k\in \mathcal{K}}\omega_k(m_l) \pi_k(l),
		\end{align}
	which is contributed from its possible load fluctuations in all non-base scenarios;

	\item generator $j$'s upward and downward reserve credit:
		\begin{equation}
		\label{upward reserve payments}
		\Gamma^U(j)=\eta^U(j) r_U(j)=\sum_{k\in \mathcal{K}}\overline{\alpha_k}^{*}(j) r_U(j)=\sum_{k\in \mathcal{K}}\Gamma^U_k(j),
		\end{equation}
		\begin{equation}
		\label{downward reserve payments}
		\Gamma^D(j)=\eta^D(j) r_D(j)=\sum_{k\in \mathcal{K}}\overline{\beta_k}^{*}(j) r_D(j)=\sum_{k\in \mathcal{K}}\Gamma^D_k(j).
		\end{equation}
	\end{itemize}
	
	\begin{remark}
	\label{Remark}
    {\rm The co-optimization model (II) is proposed to guard against all possible non-base scenarios. Although only one scenario will be realized, other ones have also contributed to the cost of procuring reserve. How to properly distribute the cost of guarding against scenarios that do not actually happen is thus a key issue. In the proposed settlement process, consumers pay for their possible fluctuations in all non-base scenarios as in (\ref{fluctuation payment}). 
    
    An alternative is letting consumers pay only for their actual fluctuations in the realized scenario. In this case, the fluctuation payment is ex-post and formulated as}
		\begin{equation}
		\label{fluctuation payment 2}
			\Gamma^{\pi}_k(l)
			=\frac{\omega_k(m_l)}{\epsilon_k} \pi_k(l),
	\end{equation}
    {\rm and the SO will afford the cost of other hypothetical scenarios. As shown in \cite{Morales2012} and in this paper, both approaches can achieve the revenue adequacy for the SO in expectation. However, we argue that load payments and merchandise surplus of the SO are much more volatile in the approach in (\ref{fluctuation payment 2}). The reason is that the reserve cost in the co-optimization model (II) mainly comes from some severe but rare scenarios. In most cases where these extreme scenarios are not realized, the load payment will be relatively low, and the net revenue of the system operator will be negative. However, if one of the extreme scenarios happens, load payment will increase significantly, leading to large amounts of SO's net revenue that can offset negative values in normal scenarios. Such drastic increases in load payment under rare but extreme scenarios, however, are sometimes unacceptable, as in the Texas power crisis in early 2021. This is an important reason for us to adopt the ex-ante settlement for possible load fluctuations. Simulations in Section VI also verify our intuitions above.}
    \end{remark}

    \subsubsection{ex-post stage}	 
	In this stage, if the base case happens, no adjustment is needed. Otherwise, assume non-base scenario $k$ happens, then the generation re-dispatch of each generator $j$ will be either $\delta g^U_k(j)$ or $\delta g^D_k(j)$, and will be settled with re-dispatch prices $\overline{c}$ and $\underline{c}$. Load shedding will be settled with shedding prices $c_L$. Therefore, the ex-post stage includes the following payments: 
	\begin{itemize}
		\item upward re-dispatch credit:
		\begin{equation}
		\label{upward redispatch payments}
		\Phi^U_k(j)=\overline{c}(j)\delta g^U_k(j);
		\end{equation}
		\item downward re-dispatch pay-back:
		\begin{equation}
		\label{downward redispatch payments}
		\Phi^D_k(j)=-\underline{c}(j)\delta g^D_k(j);
		\end{equation}
		\item load shedding compensation:
		\begin{equation}
		\label{load shedding credits}
		\Phi^d_k(l)=c_L(l)\delta d_k(l).
		\end{equation}
	\end{itemize}
	
	Note that ex-post re-dispatches and load shedding must be implemented in a very short period of time, sometimes automatically. We therefore do not introduce another round of optimizations or auctions. Re-dispatch prices and load shedding prices in (\ref{obj2}) will be used as settlement prices instead of marginal prices, which makes the ex-post settlement paid-as-bid.
	
	With the proposed model, the pricing approach and the settlement process, some attractive properties can be established, as in the next section.
	 
	\section{Properties}	 
	
	In this section, we investigate several key properties of the proposed co-optimization and pricing scheme: proportional uniform pricing for re-dispatch, individual rationality, cost recovery for generators for each scenario, and revenue adequacy for the system operator in expectation.
	
	\subsection{Proportional Uniform Pricing for Re-dispatch}
	In the proposed settlement process, the revenue of a generator consists of three parts: (i) the ex-ante energy credit in (\ref{base-case prices to generator energy credit}) and (\ref{non-base scenario prices to generator energy credit}), (ii) the ex-ante upward reserve credit in (\ref{upward reserve payments}) and downward reserve credit in (\ref{downward reserve payments}), and (iii) the ex-post upward re-dispatch credit in (\ref{upward redispatch payments}) and downward re-dispatch pay-back in (\ref{downward redispatch payments}). With Theorem 1 in Section III, the property of locational uniform pricing has been established for the generator energy credit. Next we will consider establishing this property for the second and third part of generator credit as a whole. For simplicity, we only analyze the upward reserve and re-dispatch credit in this subsection, and the analysis can be easily applied to downward reserve and re-dispatch. 
	
	Note that the generator reserve revenue in (\ref{upward reserve payments}) and the expectation of the generator re-dispatch credit in (\ref{upward redispatch payments}) are written in a scenario-wise form as follows:
	\begin{equation}
	\label{Bernoulli expect}
	\sum_k(\Gamma^U_k(j)+\epsilon_k\Phi^U_k(j))=\sum_k\Pi^U_k(j).
	\end{equation}
	Each term $\Pi^U_k(j)$ in (\ref{Bernoulli expect}) can be interpreted as the fractional contribution of scenario $k$ to the reserve and expected re-dispatch revenue of generator $j$. Next, we show that such fractional revenues of different generators in the same scenario $k$ are proportional to their re-dispatch quantities, which are referred to as the proportional locational uniform pricing property in this paper:

	\begin{theorem}[Proportional Uniform Pricing for Re-dispatch]
	For any given scenario $k$, consider any two generators $i,j$ at the same bus. Under assumptions (i)-(v) and assume that $\delta g^U_k(i),\delta g^U_k(j)>0$,
	\begin{align}
	\label{uniform pricing of redispatch}
	\frac{\Pi^U_k(i)}{\delta g^U_k(i)}=\frac{\Pi^U_k(j)}{\delta g^U_k(j)}=\omega_k(m_i)=\omega_k(m_j),\forall k \in \mathcal{K}.
	\end{align}
	\end{theorem}
    Please check the Appendix \ref{Sec:LMP reserve} for the proof. Note that neither the reserve revenue in (\ref{upward reserve payments}) nor the re-dispatch credit in (\ref{upward redispatch payments}) alone has similar uniform pricing property. Essentially, the property of uniform pricing is a result of “the law of one price”: Under certain conditions, identical goods should have the same price. However, reserve procured from different generators at the same bus may not be identical, due to their different re-dispatch costs. Such a property is only true if we consider the entire re-dispatch process (\ref{upward reserve payments}) and (\ref{upward redispatch payments}) as a whole.
	
	\subsection{Cost Recovery}
    We review properties of market participants in the proposed co-optimization model and pricing mechanism. To establish the properties in this subsection, we introduce another assumption to get rid of the non-zero lower bound issue of generators, which may lead to uplifts. The additional assumption is presented as follows:
    
    vii) The lower bound of each generator's energy output $\underline{G}$ is assumed to be zero. 
    
    First, we establish the property of individual rationality as the following:
    \begin{theorem}[Individual Rationality]
	Under assumptions (i)-(v) and (vii), consider any generator $j$. We assume that its procured quantities of energy and reserve $(g^*(j), r_U^*(j), r_D^*(j))$ are solved from model (II) and its settlement prices for energy and reserve $(\eta^g(j),\eta^U(j),\eta^D(j))$ are calculated by (\ref{RGMP2}), (\ref{RUMP}) and (\ref{RDMP}), respectively. Then
	\begin{align}
	\label{IR1}
	&g^*(j)\!,r_U^*(j)\!,r_D^*(j)\notag\\
	&=\underset{g(j),r_U(j),r_D(j)}{\operatorname{argmax}}\{F^{IV}_j(g(j)\!,r_U(j)\!,r_D(j))|(\ref{base physical limit1}),(\ref{base physical limit2})\},
	\end{align}
	where 
	\begin{align}
	\label{IR2}
	F^{IV}_j(g(j)\!,r_U(j)\!,r_D(j))\!&=\!\eta^g(j)g(j)\! +\! \eta^U\!(j)r_U(j) \!+ \! \eta^D\!(j)r_D(j)\!\notag\\
	 &-\!c_{g}(j)g(j) \!- \!c_{U}(j)r_U(j) \!-\! c_{D}(j)r_D(j).
	\end{align}
	\end{theorem}
	In other words, if generator $j$ were able to freely adjust its supply of energy and reserve with given prices $(\eta^g(j),\eta^U(j),\eta^D(j))$, then the solution $(g^*(j), r_U^*(j), r_D^*(j))$ to the co-optimization model (II) would have been maximized its profit.

	Please check the Appendix \ref{Sec:Individual Rationality} for the proof. Although Theorem 3 is established for the ex-ante stage only, it is still valid considering the ex-post stage. This is because the ex-post settlement is true-cost based and will not affect the profit of generators. 
	
	A natural corollary of Theorem 3 is the property of cost recovery for generators:

	\begin{corollary}[Cost Recovery]
	Under assumptions (i)-(v) and (vii), with the realization of any scenario, the total credit of any generator $j$ is no less than its total bid-in cost of energy, reserve, and re-dispatch, i.e.,
    \begin{align}
	&\eta^g(j)g(j)\! +\! \eta^U\!(j)r_U(j) \!+ \! \eta^D\!(j)r_D(j) \notag\\
	\geq &c_{g}(j)g(j) \!+ \!c_{U}(j)r_U(j) \!+\! c_{D}(j)r_D(j).
	\end{align}
	\end{corollary}

    The properties of individual rationality and cost recovery for generators are established from the perspective of market participants. Next we will take the system operator's point of view and establish the revenue adequacy property.
	
	\subsection{Revenue Adequacy}
	For the system operator, its total congestion rent $\Delta$ from the proposed model (II) will be the sum of the contribution $\Delta_0$ from the base case and the contributions $\sum_{k\in \mathcal{K}}\Delta_k$ form all non-base scenarios. $\Delta_0$ and $\Delta_k$ are calculated by the following equations:
	\begin{align}
	\label{congestion rent basecase}
	\Delta_0=f^T\mu,
	\end{align}	
	\begin{align}
	\label{congestion rent scenario k}
	\Delta_k=f^T_k\mu_k.
	\end{align}
	For the proposed settlement process, the following property regarding the SO's revenue adequacy can be established: 
	\begin{theorem}[Revenue Adequacy]   
    Under assumptions (i)-(vi), in expectation, total load payment is equal to the sum of total generator credit and total congestion rent $\Delta$.

	Moreover, the property of revenue adequacy can be decomposed scenario-wise. Namely, the base-case load energy payment (\ref{base-case prices to load payment}) is equal to the sum of the base-case generator energy credit (\ref{base-case prices to generator energy credit}) and the base-case congestion rent (\ref{congestion rent basecase}):
	\begin{equation}
	\mathds{1}^T\Gamma^d_0=\mathds{1}^T\Gamma^g_0+\Delta_0.
	\end{equation}
	And for each non-base scenario $k$, the sum of its contribution to load payment $\mathds{1}^T(\Gamma^d_k \!+\!\Gamma^{\pi}_k)$ in (\ref{non-base scenario prices to load payment in all scenarios})-(\ref{fluctuation payment}) is equal to the sum of its contribution to generator ex-ante energy credit $\mathds{1}^T\Gamma^g_k$ in (\ref{non-base scenario prices to generator energy credit}) and reserve credit $\mathds{1}^T(\Gamma^U_k\!+\!\Gamma^D_k)$ in (\ref{upward reserve payments})-(\ref{downward reserve payments}), the expected generator ex-post re-dispatch payment $\mathds{1}^T(\epsilon_k\Phi^U_k \!+\!\epsilon_k\Phi^D_k)$ in (\ref{upward redispatch payments})-(\ref{downward redispatch payments}), the expected load shedding compensation $\mathds{1}^T(\epsilon_k\Phi^d_k)$ in (\ref{load shedding credits}) , and the congestion rent $\Delta_k$ in (\ref{congestion rent scenario k}):
	\begin{align}
    \label{RA Theorem}
	&\mathds{1}^T(\Gamma^d_k +\Gamma^{\pi}_k)\notag\\
	=&\mathds{1}^T(\Gamma^g_k+\Gamma^U_k+\Gamma^D_k+\epsilon_k\Phi^U_k +\epsilon_k\Phi^D_k+\epsilon_k\Phi^d_k)+\Delta_k,\forall k \in \mathcal{K}.
	\end{align} 
	\end{theorem}
	
    Please refer to the Appendix \ref{Sec:Revenue Adequacy} for the proof. With (\ref{RA Theorem}), it can be observed that in expectation, ex-ante reserve procurement cost and ex-post re-adjustment cost will be allocated scenario-wise to loads based their load fluctuation severity.

    Some existing works also consider the scenario-oriented model and pricing scheme, see \cite{Morales2012} or \cite{readjust1}. Regarding pricing approaches, there is no absolute right or wrong. We argue that the proposed method is attractive because (i) it strictly follows the marginal pricing principle, and (ii) many important properties have been established.
    
    \section{Discussions}
    In this section, we will revisit assumptions (ii)-(iv) and discuss their implications.
    
    In assumption (ii), we assume that a single-period problem is considered for simplicity, whereas practical reserve markets may follow a multi-period setting. The biggest challenge in a multi-period problem is the coupling between reserve and ramping, which is a highly complicated issue, see \cite{MultiInterval} as an example. Since the main purpose of this paper is to focus on the reserve issue, studying properties of the co-optimization and associated market mechanism, we will leave the multi-period problem to our future works.
    
    In assumption (iii), we assume that renewable generations are modelled as negative loads. In essence, we assume that the system always accommodates all renewable energy available in the ex-ante energy procurement. Consequently, renewable generators will have the same effects as loads from the perspective of the co-optimization model (II). 
    
   In assumption (iv), we ignore generator outages in non-base scenarios to establish the property of locational uniform pricing for energy. If generator outages are considered, the co-optimization model and prices can be obtained similarly. In particular, the objective function (\ref{obj2}) and base-case constraints (\ref{base balance and pf})-(\ref{base physical limit2}) will remain the same, while non-base constraints (\ref{cntg balance})-(\ref{dg2fanwei}) will be modified as
\begin{align}
\label{cntg balance 1}
&\lambda_k : 1^T(\widetilde{g}_k+\delta g^U_k-\delta g^D_k)=1^T (d+\pi_k -\delta d_k),\\
\label{cntg pf 1}
&\mu_k : S_{G,k}^T(\widetilde{g}_k+\delta g^U_k-\delta g^D_k)-S_{D,k}^T(d+\pi_k -\delta d_k) \leq f_k,\\
\label{dg1fanwei 1}
&(\underline{\alpha_k},\overline{\alpha_k}): 0 \leq \delta g^U_k \leq \widetilde{r}_{U,k},\\
\label{dg2fanwei 1}	
&(\underline{\beta_k},\overline{\beta_k}) : 0 \leq \delta g^D_k \leq \widetilde{r}_{D,k},
\end{align}    
where $\widetilde{g}_k,\widetilde{r}_{U,k},\widetilde{r}_{D,k}$ are the vectors of available generations and reserve capacities in scenario $k$ considering possible generator outages in that scenario. With these modifications, the proposed model can efficiently model generator outages and optimally procure reserve to guard against them. 

Moreover, considering the prices of energy and reserve, for generator $j$, its energy and reserve prices in (\ref{RGMP2}) and (\ref{RUMP})-(\ref{RDMP}) will be modified as

	\begin{align}
	\label{RGMP 3}
	\eta^g(j)&  = -\frac{\partial F_{-j}(x_{-j}^*)}{\partial g(j)}\!\notag\\
	&=\!\lambda^*\!-\!S_G(\cdot,m_j)^T\!\mu^*\!+\!\sum_{k\in \mathcal{K},j \not\in \Omega_k}\!(\lambda_k^*\!-\!S_{G,k}(\cdot,m_j)^T\!\mu_k^*)\notag\\
	&=\omega_0(j)+\sum_{k\in \mathcal{K},j \not\in \Omega_k}\omega_k(j),\\
	\label{RUMP 2}
	\eta^U(j) &  = -\frac{\partial F_{-j}(x_{-j}^*)}{\partial r_U(j)}=\sum_{k\in \mathcal{K},j \not\in \Omega_k} \overline{\alpha_k}^*(j),\\
	\label{RDMP 2}
	\eta^D(j) &  = -\frac{\partial F_{-j}(x_{-j}^*)}{\partial r_D(j)}=\sum_{k\in \mathcal{K},j \not\in \Omega_k} \overline{\beta_k}^*(j),
	\end{align}
where $\Omega_k$ is the set of generators that are shut down in scenario $k$. Compared with the original price formulations in (\ref{RGMP2}) and (\ref{RUMP})-(\ref{RDMP}), it can observed that if generator $j$ is shut down in scenario $k$, then this scenario should be excluded from the calculation of energy and reserve prices in (\ref{RGMP 3})-(\ref{RDMP 2}).

With the energy and reserve prices in (\ref{RGMP 3})-(\ref{RDMP 2}), we also establish revenue adequacy for the system operator in expectation with the similar procedure in the Appendix \ref{Sec:Revenue Adequacy}. However, considering generator outages will bring in the complicated issue of non-uniform pricing. For energy, considering different generators at the same bus with different outage probabilities, they will receive different energy marginal prices because their generations are no longer homogeneous goods. In addition, for reserve, considering different generators at the same bus, even if they have the same upward and downward re-dispatch price, their reserve marginal prices will be different if they have different outage probabilities. Furthermore, other counter-intuitive results have been found in our simulations. For example, for two generators at the same bus with the same outage probability, the generator with higher output will receive lower energy marginal price because its possible outage is more severe. These complicated issues cannot be fully addressed as a part of this paper, therefore more efforts will be made to digest these results in our future studies.

	\section{Case Study}
    Case studies were performed both on a 2-bus system and modified IEEE 118-bus system.
	\subsection{Two-Bus System}
	 \begin{figure}[ht]
		\centering
		\captionsetup{font=footnotesize,singlelinecheck=false}
		\includegraphics[width=3in]{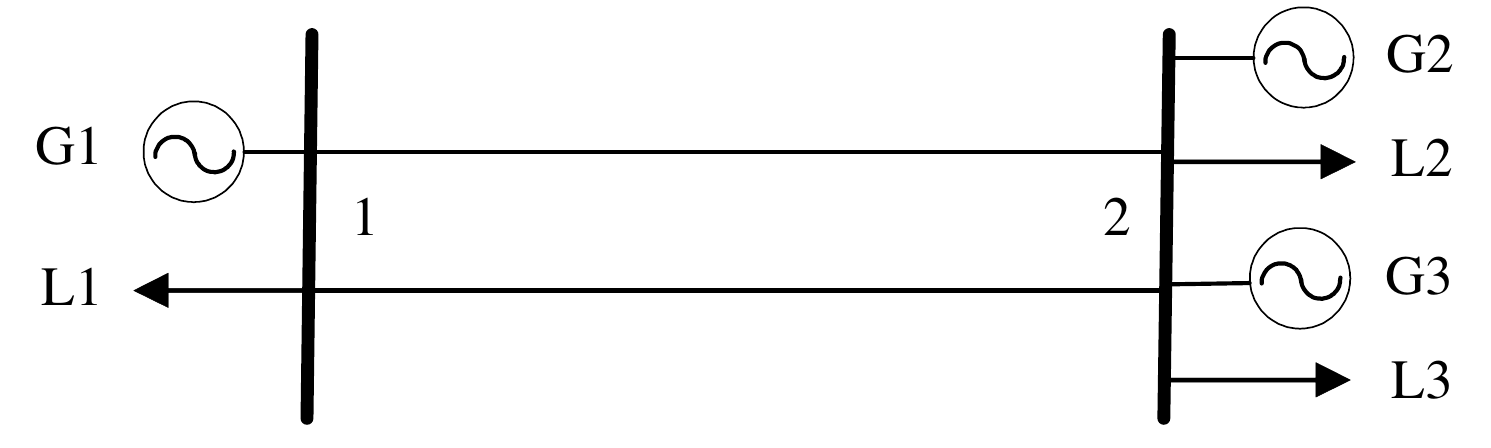}
		\caption{One-Line Diagram of the 2-Bus System}
		\label{one line diag}
	    \vspace{-0.2cm}	
	\end{figure}
    First, a simple two-bus system was adopted to illustrate the proposed co-optimization and pricing mechanism, as well as properties thereof. The one-line diagram was presented in Fig. \ref{one line diag}. There are two parallel and identical transmission lines, each with a capacity of 1MW. For non-base scenarios, one of the two parallel lines may be cut off. In addition, in non-base scenarios, the power flow limit on each transmission line is set as 1.2MW. Furthermore, generator parameters of this 2-bus case were presented in Table \ref{Generators' Offers}, and the base load vector for L1, L2 and L3 is (6,15,4)MW. In addition, all possible non-base scenarios for this 2-bus case were given in Table \ref{Non-base}, and the probability of the base case happening is $1-\sum \epsilon_k=0.54$.
    
    	\begin{table}[ht]
	    \footnotesize
		\centering
		\vspace{-0.2cm}
		\setlength{\abovecaptionskip}{-0.5pt}
        \setlength{\belowcaptionskip}{0pt}
        \captionsetup{font=footnotesize,textfont=sc}
		\caption{Generator Parameters of the 2-Bus Case}
		\begin{tabular}{cccc}
			\toprule
			Generator &$\overline{G}/\underline{G}$& $\overline{r_U}/\overline{r_D}$& $c_{g}/c_{U}/c_{D}$ \\
			\hline
		 G1  & $16/0$ & $4/4$ & $8/2/2$\\
		 G2  &$18/0$  &  $4/4$ & $15/2/2$  \\
		 G3 & $12/0$  &  $4/4$ &  $20/2.5/2.5$   \\
			\bottomrule
			\end{tabular}
		\label{Generators' Offers}
		\end{table}
		
	\begin{table}[ht]
	\centering
	\vspace{-0.3cm}
	\footnotesize
    \setlength{\abovecaptionskip}{-0.5pt}
    \setlength{\belowcaptionskip}{0pt}
        \captionsetup{font=footnotesize,textfont=sc}
			\caption{Non-Base Scenarios for the 2-Bus Case}
			\label{Non-base}
			\begin{tabular}{cccc}
				\toprule
				NO.  & Line outage & Load & Probability \\
				\hline
				1 & Yes & (6,15,4) & $0.06$ \\
				2/3 & Yes & (8,21,3)/(9,17,1)  & $0.02$/$0.02$  \\
				4/5 & No & (8,21,3)/(9,17,1)  &$0.18$/$0.18$ \\
				\bottomrule
			\end{tabular}
		\vspace{-0.3cm}
		\end{table}
		
	Market clearing results and prices of the 2-bus case with the proposed model (II) were presented in Table \ref{clear}. For cleared quantities, note that although G1 offers the cheapest upward reserve and still has extra generation capacity and ramping rate, the SO does not clear G1's entire upward reserve. Instead, the more expensive upward reserve resources G2 and G3 are cleared. The reason is that the extra upward reserve from G1 will not be deliverable in scenarios with line outages. In addition, from the $5^{th}$ column of Table \ref{clear}, it can be confirmed that the energy prices are locational uniform. Furthermore, from the $6^{th}-7^{th}$ columns it can be observed that G2 receives higher upward reserve price but lower downward reserve price compared with G3 because G2's upward and downward re-dispatch prices $\overline{c},\underline{c}$ are both lower.

		\begin{table}[ht]
		\footnotesize
			\centering
        \vspace{-0.2cm} 
        \setlength{\abovecaptionskip}{-0.5pt}
        \setlength{\belowcaptionskip}{0pt}
            \captionsetup{font=footnotesize,textfont=sc}
			\caption{Clearing Results and Prices of the 2-Bus Case}	
			\label{clear}
			\begin{tabular}{ccccccc}
				\toprule
				Generator & $g$ & $r_U$ & $r_D$& $\eta^g$ & $\eta^U$ & $\eta^D$\\
				\hline
				G1 & $8.0$ & $2.4$ & $0.8$& $8.0$ & $2.0$ & $2.0$ \\
				G2 & $16.4$ & $1.6$ & $0.0$& $20.0$ & $7.0$ & $2.0$ \\
				G3 & $0.6$ & $4.0$ & $0.4$& $20.0$ & $6.0$ & $2.5$ \\
				\bottomrule
			\end{tabular}
		\vspace{-0.2cm}
		\end{table}
	
	Moreover, the proportional locational uniform pricing property for re-dispatch was illustrated in Table \ref{uniform}. We have $\delta g^U_k(2),\delta g^U_k(3)>0$ for G2 and G3 in scenario 2 and 4, and in these two scenarios we have	$\frac{\Pi^U_k(2)}{\delta g^U_k(2)}\!=\!\frac{\Pi^U_k(3)}{\delta g^U_k(3)}\!=\!\omega_k(m_2)\!=\!\omega_k(m_3)$, validating Theorem 2.
	
	\begin{table}[H]
		\footnotesize
			\centering
        \vspace{-0.2cm} 
            \captionsetup{font=footnotesize,textfont=sc}
			\caption{Proportional Locational Uniform Pricing for Re-dispatch in the 2-bus Case}	
			\label{uniform}
			\begin{tabular}{cccccc}
				\toprule
				Scenario & Generator &$\delta g^U_k$ &$\Pi^U_k$ &$\frac{\Pi^U_k}{\delta g^U_k}$ &$\omega_k$\\
				\hline
				\multirow{2}{*}{Scenario 2} & G2& $1.60$& $1.92$& $1.20$& $1.20$\\ 
				& G3& $4.00$& $4.80$& $1.20$& $1.20$\\
				\hline
				\multirow{2}{*}{Scenario 4} & G2& $1.60$& $17.08$& $8.80$& $8.80$\\ 
				& G3& $4.00$& $35.20$& $8.80$& $8.80$\\
				\bottomrule
			\end{tabular}
		\vspace{-0.2cm}
		\end{table}		
			
	Furthermore, with the realization of any non-base scenario, profits of G1, G2 and G3 are $\$101.60,\$298.28,\$21.62$, respectively, which confirms the property of cost recovery. In addition, in Table \ref{money flow}, the money flow in this 2-bus case was presented. We can see that in expectation, payments from loads, credits to generators and congestion rent can reach their balance for the base case as in the $2^{nd}$ column, for each scenario as in the $3^{rd}-7^{th}$ columns, and in total as in the last column, validating the property of revenue adequacy.
	
		\begin{table}[H]
		\footnotesize
		\setlength{\abovecaptionskip}{-0.5pt}
        \setlength{\belowcaptionskip}{-1.5pt}
            \captionsetup{font=footnotesize,textfont=sc}         
            \caption{Money Flow in the 2-Bus Case(\$)}
			\label{money flow}
			\centering
			\begin{tabular}{ccccccccc}
				\toprule
				& Base & S1 & S2 & S3 & S4 & S5& Total\\
				\hline
				$\Gamma^d$ & $482.0$ & $8.0$ & $23.8$ &$6.7$ & $187.4$  & $37.4$ & $745.2$\\
				$\Gamma^{\pi}$\ & $0$ &  $0$ & $7.5$ & $0.2$ & $59.5$   & $3.0$& $70.2$\\
				\hline
				$\Gamma^g$ & $476.0$& $3.2$ &  $21.7$  & $6.4$ & $176.5$  & $37.4$& $721.2$\\	
				$\Gamma^U$ & $0$ & $0$ & $4.6$ & $0$  & $35.2$  & $0.1$& $40.0$\\	
				$\Gamma^D$& $0$ & $1.7$ & $0$ & $0.1$ & $0$ & $0.8$ & $2.6$\\
				$\epsilon\Phi^U$ & $0$ & $1.1$ &$2.3$ &$0.4$ & $22.2$ & $3.5$ & $29.5$\\
				$\epsilon\Phi^D$ & $0$ & $-0.8$ & $-0.1$ & $-0.2$ & $0$ & $-1.4$ & $-2.5$\\
				$\epsilon\Phi^d$ & $0$ & $0$ & $1.4$  & $0$ & $0$  & $0$& $1.4$\\
				$\Delta$ & $6.0$& $2.9$ &$1.2$ &$0.2$& $13.0$ & $0$ & $23.3$\\
				\bottomrule
			\end{tabular}
		\end{table}

	\subsection{IEEE 118-Bus System}
	\indent Simulations on modified IEEE 118-bus system were also recorded. The transmission capacities are modified as 1.5 times of the DCOPF results of the original IEEE 118-bus system. In addition, transmission capacities that are smaller than 10MW will be set as 10MW. In non-base scenarios, power flow limits on transmission lines are set as 1.3 times of the base-case values. The generation cost of each generator is modified as linear term, and its upward and downward reserve bid-in price are set as 1/5 of its energy bid-in price. Moreover, the upward and downward ramping rate of each generator is set as 0.1 times of its generation capacity upper bound. In addition, the original load 59 is equally separated into two loads: new load 59 and load 119. Furthermore, minor modifications are also applied to some load capacities. We uploaded our Matlab case file onto Github as in \cite{modified118}. In addition, all possible non-base scenarios in this case were given in Table \ref{Non-base-118}.
	
	\begin{table}[H]
	\centering
	\footnotesize
    \setlength{\abovecaptionskip}{-0.5pt}
    \setlength{\belowcaptionskip}{0pt}
        \captionsetup{font=footnotesize,textfont=sc}
			\caption{Non-Base Scenarios for the 118-Bus Case}
			\label{Non-base-118}
			\begin{tabular}{cccc}
				\toprule
				NO.  & Outage & Load Situation & Probability \\
				\hline
				1 & No outage & d119 $\uparrow$ by 3\%, others $\downarrow$ by 3\% & $0.07$ \\
				2 & No outage & d119 $\downarrow$ by 3\%, others $\uparrow$ by 3\% & $0.07$ \\
				3 & Line 21 & d119 $\uparrow$ by 3\%, others $\downarrow$ by 3\% &$0.01$ \\
				4 & Line 21 & d119 $\downarrow$ by 3\%, others $\uparrow$ by 3\% &$0.01$ \\
				5 & Line 21 & basic load & $0.08$\\
				6 & Line 55 & d119 $\uparrow$ by 3\%, others $\downarrow$ by 3\% &$0.01$ \\
				7 & Line 55 & d119 $\downarrow$ by 3\%, others $\uparrow$ by 3\% &$0.01$ \\
				8 & Line 55 & basic load & $0.08$\\
				9 & Line 102 & d119 $\downarrow$ by 3\%, others $\uparrow$ by 3\% &$0.01$ \\
				10 & Line 102 & d119 $\downarrow$ by 3\%, others $\uparrow$ by 3\% &$0.01$ \\
				11 & Line 102 & basic load & $0.08$\\
				\hline
			\end{tabular}
		\vspace{-0.3cm}
		\end{table}
		
\begin{figure}[t]
		\centering
		\includegraphics[width=3.2in]{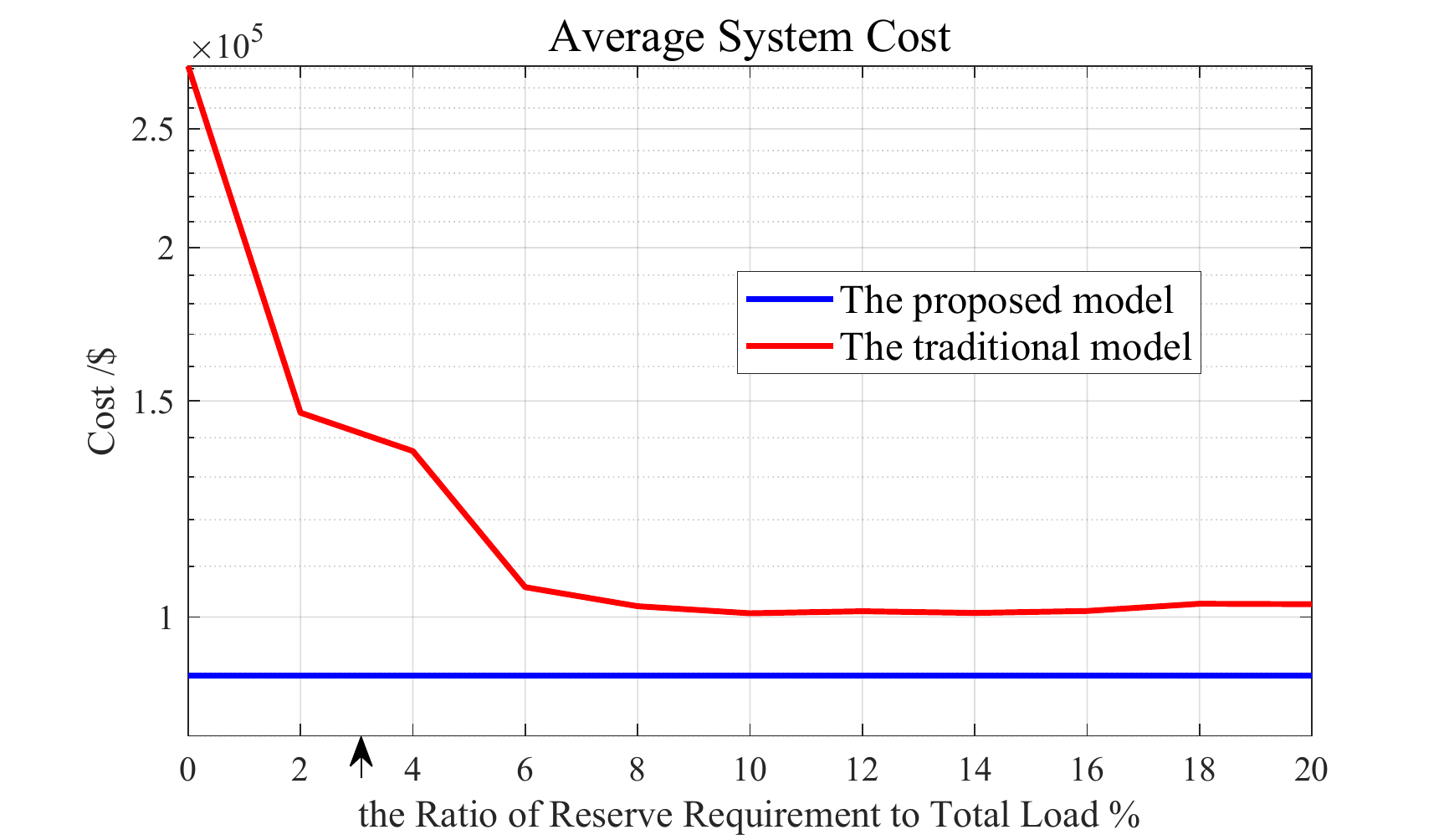}
		\captionsetup{font=footnotesize,singlelinecheck=false}
		\caption{Average system costs from the proposed model (blue) and the traditional model under different reserve requirement settings (red)}
		\label{pro_vs_tra}
	\end{figure}

	In Fig. \ref{pro_vs_tra}, with Monte Carlo Simulation, the average system cost of the proposed model was compared with that of the traditional model under different reserve requirement settings, with the following steps:
    
    (i) We selected different reserve requirements as different ratios of the system total load for the traditional model;
    
    (ii) We calculated the energy and reserve clearing results and base-case procurement costs of the modified 118-bus case with the traditional model under these different reserve requirement settings from step (i);
    
    (iii) We generated 50000 Monte Carlo Samples based on the occurrence probabilities of non-base scenarios in Table \ref{Non-base-118};
    
    (iv) We calculated the average re-adjustment costs of the traditional model under different reserve requirement settings in all Monte Carlo Samples. If in one Monte Carlo Sample, the re-adjustment problem is infeasible, then the re-adjustment cost in that case will be set as 20000, while the expected system total cost from the proposed model (II) is 89651.6. 
    
    (v) We obtained the average system costs of the traditional model under different reserve requirement settings by adding the base-case procurement costs from step (ii) to the average re-adjustment costs from step (iv), and presented them in Fig. \ref{pro_vs_tra} as the red curve. Note that with the increasing reserve requirement, the red curve first goes down because of the decreasing load shedding cost, and then goes up because of the increasing reserve procurement cost. 
    
    (vi) We repeated steps (ii)-(v) for the proposed model (II), and presented the average system cost of the proposed model in Fig. \ref{pro_vs_tra} as the blue curve. On top of that, the overall upward and downward reserve procurement of the modified 118-bus case with the proposed model $\mathds{1}^T r_U^*$ and $\mathds{1}^T r_D^*$ are both about 3\% of system total load, indicated by an arrow in Fig. \ref{pro_vs_tra}. It can be observed that the proposed model can efficiently reduce reserve procurement, and can reduce the average system cost by $10.99\%$-$68.14\%$ in this test compared with the traditional model.
	
	\begin{figure}[t]
		\centering
		\includegraphics[width=3.5in]{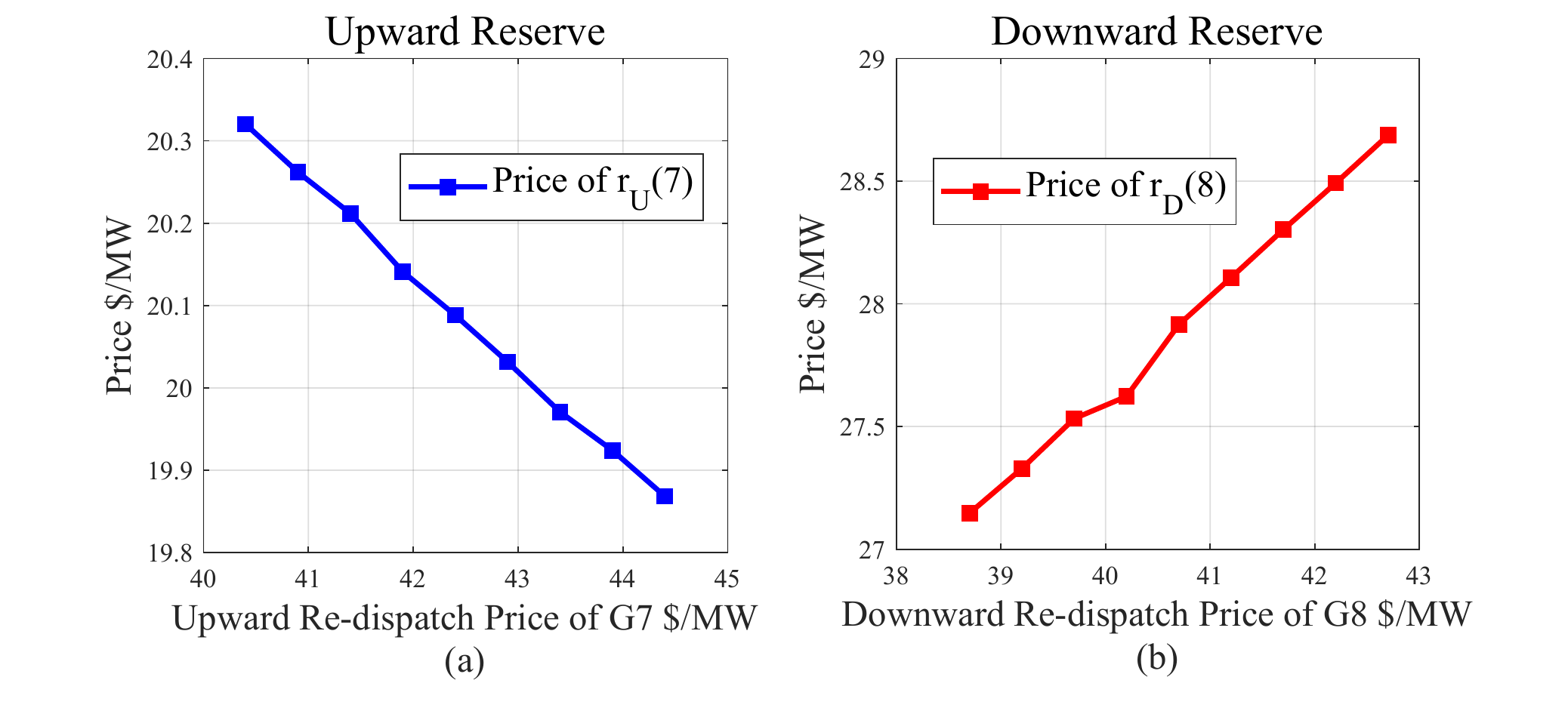}
		\vspace{-0.7cm}
		\captionsetup{font=footnotesize,singlelinecheck=false}
		\caption{(a) Price of G7's upward reserve $r_U(7)$ with respect to the increasing upward re-dispatch price of G7; (b) Price of G8's downward reserve $r_D(8)$ with respect to the increasing downward re-dispatch price of G8}
		\label{plot_cplus_cminus_on_u_b}
		\vspace{-0.2cm}
	\end{figure}
	
    In addition, in Fig. \ref{plot_cplus_cminus_on_u_b}, the relationships between reserve marginal prices and re-dispatch prices were presented. In Fig. \ref{plot_cplus_cminus_on_u_b}(a), with the increasing upward re-dispatch price of G7, the upward reserve price of G7 will decrease. At the same time, In Fig. \ref{plot_cplus_cminus_on_u_b}(b), with the increasing downward re-dispatch price of G8, the downward reserve price of G8 will increase because of the negative sign before the term $\underline{c}^T \delta g^D_k$ in (\ref{obj2}).

		\begin{figure}[t]
		\centering
		\includegraphics[width=3.55in]{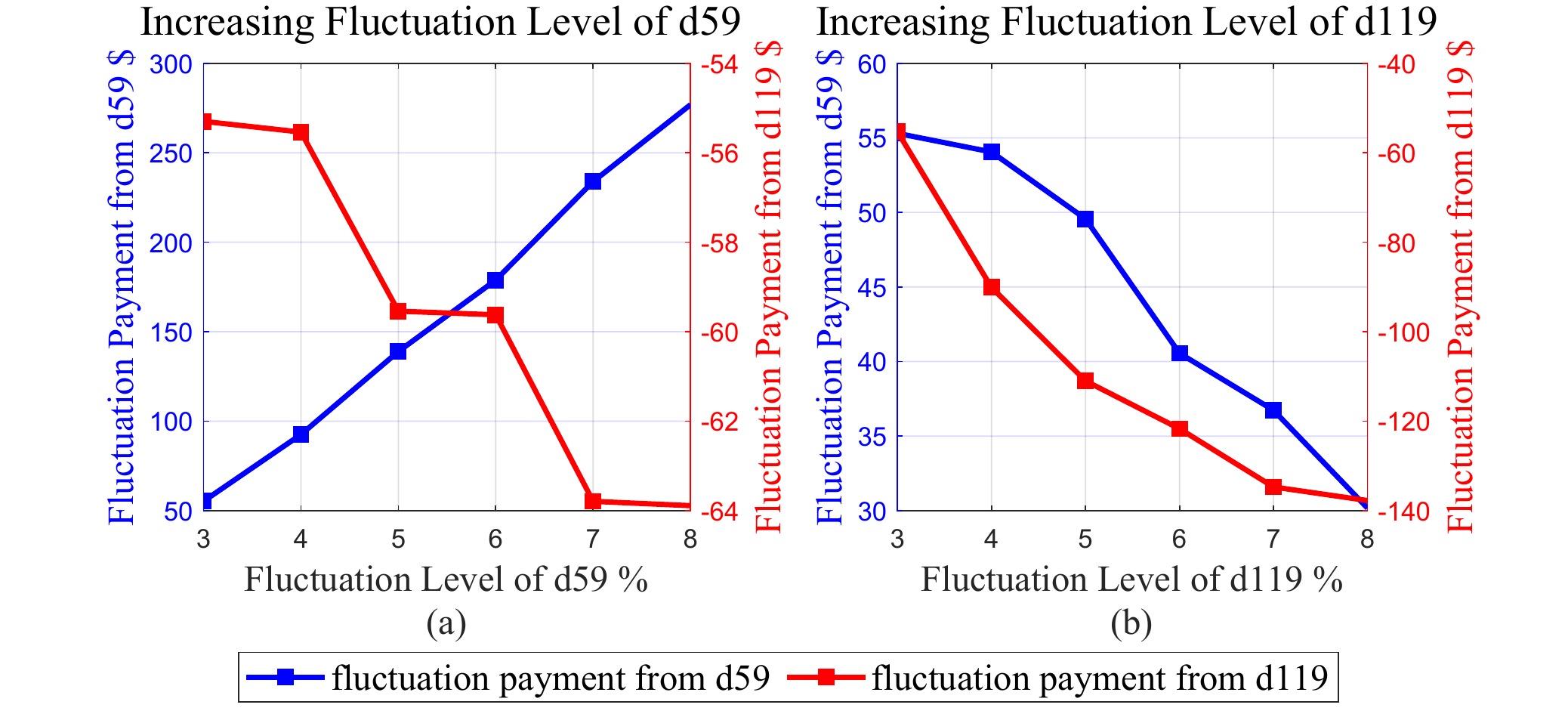}
		\vspace{-0.25cm}
        \setlength{\abovecaptionskip}{-1.5pt}
        \setlength{\belowcaptionskip}{-3.5pt}
	    \captionsetup{font=footnotesize,singlelinecheck=false,}
		\caption{(a) Fluctuation payments from d59 (blue) and d119 (red) with respect to the increasing fluctuation level of d59; (b) Fluctuation payments from d59 (blue) and d119 (red) with respect to the increasing fluctuation level of d119}
		\label{plot_M_on_fluc_d_1_2}
	\end{figure}

    In addition, from Table \ref{Non-base-118}, it can be observed that in non-base scenarios with load fluctuations, the fluctuation levels of all loads are 3\%. We fixed the fluctuation levels of all other loads except for d59 and d119, and showed how the fluctuation payments from d59 and d119 in (\ref{fluctuation payment}) changed with the increasing fluctuation level of d59 in Fig. \ref{plot_M_on_fluc_d_1_2}(a), and how they changed with the increasing fluctuation level of d119 in Fig. \ref{plot_M_on_fluc_d_1_2}(b), where the blue curve represents the fluctuation payment from d59, and the red curve represents the fluctuation payment from d119. In both Fig. \ref{plot_M_on_fluc_d_1_2}(a) and Fig. \ref{plot_M_on_fluc_d_1_2}(b), the fluctuation payment from d119 is always negative because d119's fluctuation can offset the fluctuations of other loads in non-base scenarios as shown in Table \ref{Non-base-118}. In Fig. \ref{plot_M_on_fluc_d_1_2}(a), with the rising fluctuation level of d59, both the fluctuation payment from d59 and the fluctuation credit to d119 increase: For d59, its rising fluctuation level brings in more uncertainties to the system; At the same time, for d119, considering the increasing uncertainty brought by d59, the value of its possible load fluctuation has become higher, so the fluctuation credit to d119 increases. In Fig. \ref{plot_M_on_fluc_d_1_2}(b), while the fluctuation credit to d119 increases with its rising fluctuation level because its rising fluctuation level can enhance the offset, note that the fluctuation payment from d59 decreases because the rising fluctuation level of d119 can reduce the impact of d59's fluctuation on system balance.
    
		\begin{figure}[t]
		\centering
		\includegraphics[width=3.7in]{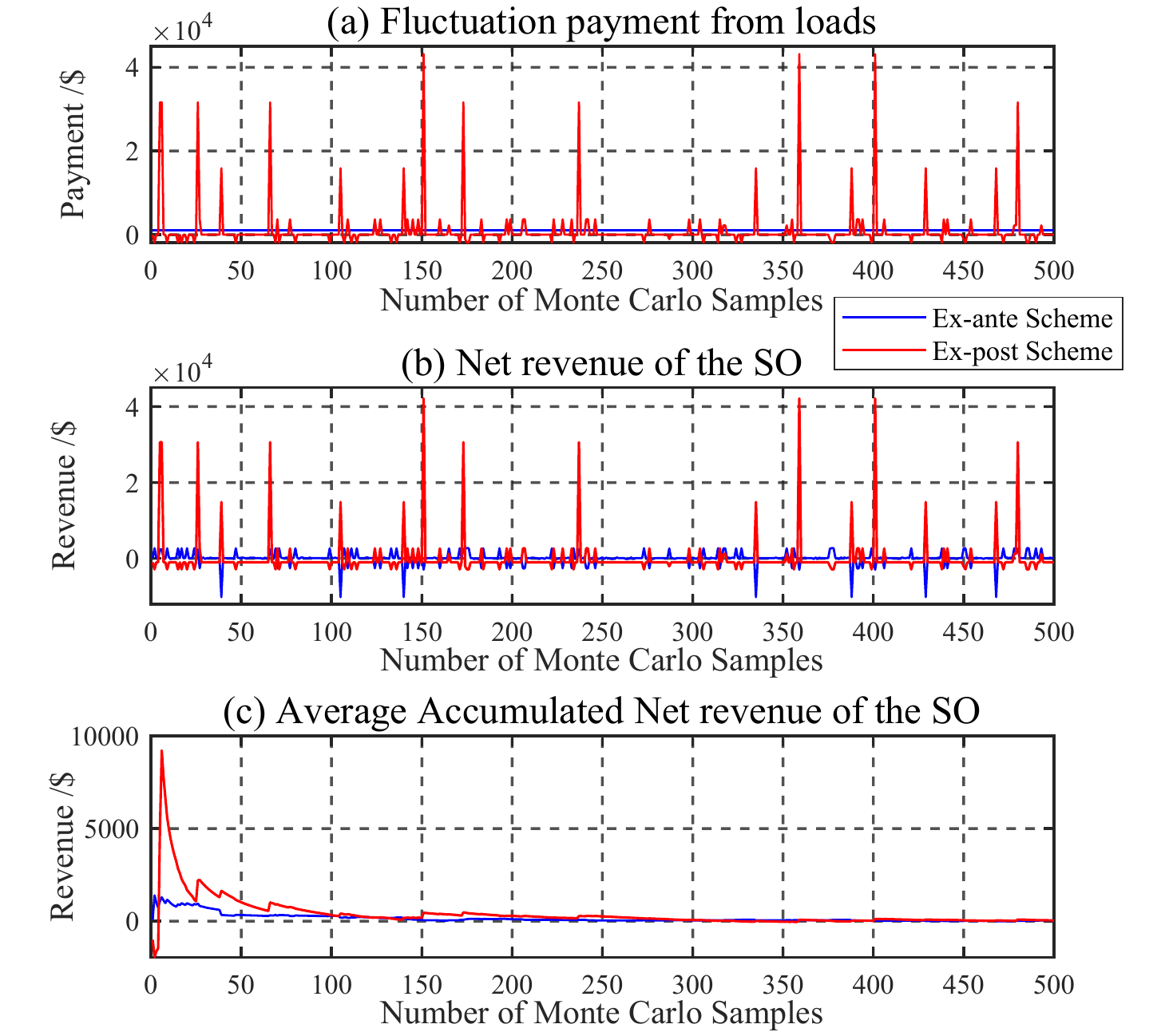}
		\vspace{0.25cm}
        \setlength{\abovecaptionskip}{-8.5pt}
        \setlength{\belowcaptionskip}{-5.5pt}
	    \captionsetup{font=footnotesize,singlelinecheck=false,}
		\caption{(a) Fluctuation payment from loads in much Monte Carlo Samples under the ex-ante scheme (blue) and the ex-post scheme (red); (b) Net revenue of the SO in much Monte Carlo Samples under the ex-ante scheme (blue) and the ex-post scheme (red); (c) Average accumulated net revenue of the SO in much Monte Carlo Samples under the ex-ante scheme (blue) and the ex-post scheme (red)}
		\label{Fig:net_revenue}
		\vspace{-0.3cm}
	\end{figure}
	
    Furthermore, to validate what we have discussed in Remark \ref{Remark}, the fluctuation payment from loads and the net revenue of the SO under the ex-ante scheme in (\ref{fluctuation payment}) are compared with that under the ex-post scheme in (\ref{fluctuation payment 2}), with the following steps:
    
    (i) We generated 500 Monte Carlo Samples based on the occurrence probabilities of non-base scenarios in Table \ref{Non-base-118};
    
   (ii) We calculated the fluctuation payment in each Monte Carlo Sample under the ex-post scheme, and presented it with the red curve in Fig. \ref{Fig:net_revenue}(a). At the same time, under the ex-ante scheme, the fluctuation payment is fixed at \$1081.67 in all Monte Carlo Samples, and we presented it with the blue curve in Fig. \ref{Fig:net_revenue}(a);
   
   (iii) We calculated the net revenue of the SO in each Monte Carlo Sample under the ex-post scheme, and presented it with the red curve in Fig. \ref{Fig:net_revenue}(b). At the same time, under the ex-ante scheme, we calculated the net revenue of the SO in each Monte Carlo Sample and presented it with the blue curve in Fig. \ref{Fig:net_revenue}(b);
   
   (iv) We calculated the average accumulated net revenue of the SO under the ex-post scheme among different numbers of Monte Carlo samples from 1, 2, ... to 500, and presented it with the red curve in Fig. \ref{Fig:net_revenue}(c). At the same time, under the ex-ante scheme, we calculated the average accumulated net revenue of the SO and presented it with the blue curve in Fig. \ref{Fig:net_revenue}(c). It can be observed that under the ex-ante scheme, the fluctuation payment from loads is fixed at \$1081.67. On the contrary, under the ex-post scheme, the fluctuation payment from loads are extremely high in some Monte Carlo Samples where one of the extreme scenario happens. For example, in the $151^{st}$ Monte Carlo Sample, the fluctuation payment under the ex-post scheme is \$43178.64, which is about 40 times as much as the fluctuation payment under the ex-ante scheme. This reveals the financial risk of consumers under the ex-post scheme. In addition, from Fig. \ref{Fig:net_revenue}(b), it can be observed that under the ex-ante scheme, the net revenue of the SO is very closed to zero in most Monte Carlo Samples. On the contrary, under the ex-post scheme, the net revenue of the SO is negative in most Monte Carlo Samples. In those Monte Carlo Samples where one of the extreme scenario happens and the fluctuation payment is large, the SO will earn much money. This reveals the financial risk of the SO under the ex-post scheme. Moreover, from Fig. \ref{Fig:net_revenue}(c), it can be observed that with the increasing number of Monte Carlo Samples, the average accumulated net revenue of the SO will converge to 0 under both the ex-ante scheme and the ex-post scheme, but the convergence speed under the ex-ante scheme will be faster. 
   
   Furthermore, considering more numbers of Monte Carlo samples, i.e., 50000, the average accumulated net revenue of the SO under the ex-ante scheme is \$6.00, which is extremely small considering the expected system total cost of this modified 118-bus case \$89648.5. Therefore, we can further confirm the property of revenue adequacy.

	\section{Conclusions}
	
	Traditional energy-reserve co-optimization highly relies on empirical reserve zones and zonal reserve requirements. In this paper, a scenario-oriented energy-reserve co-optimization model is developed, considering congestions and re-adjustment costs of all non-base scenarios. As a result, reserve resources can be optimally procured system-wide to guard against possible contingencies and load/renewable fluctuations.
	
	In addition, prices of energy and reserve have been derived based on their marginal costs/benefits to the joint clearing of energy and reserve. A key question is that should energy, reserve, and re-dispatch at the same bus be considered as homogeneous goods. If they are, under given assumptions, they will be settled at uniform prices. We have also established properties of cost recovery for generators and revenue adequacy for the system operator.
	
	In future studies, more efforts will be made to the coupling of reserve and ramping in multi-period operation and the modelling and pricing of generator outages.

	\ifCLASSOPTIONcaptionsoff
	\newpage
	\fi

	
	
	%
	%
	%
    \bibliographystyle{ieeetr}    
	\bibliography{lib_test}	
	
	\appendices
	\renewcommand{\theequation}{\arabic{equation}}

    \section{Proof of Theorem 2 (Proportional Locational Uniform Pricing for Re-dispatch)}
    \label{Sec:LMP reserve}
    According to the KKT conditions and ignoring $\overline{\tau_k}(l)$ as mentioned in assumption (vi), we have: 
	\begin{align}
	\label{L/a}
	&\frac{\partial \mathcal{L}_{\uppercase\expandafter{\romannumeral2}}}{\partial \delta d_k(l)}\!=\!
	 \epsilon_k c_{L}(l) \!-\!\underline{\tau_k}^*(l)
	 \!-\!\lambda_k^* \!+ \! S_{D,k}(\cdot,m_l)^T\mu_k^*= 0,
	\end{align}
	\begin{align}
	\label{L/b}
	\frac{\partial \mathcal{L}_{\uppercase\expandafter{\romannumeral2}}}{\partial \delta g^U_k(j)} &= 
    \epsilon_k\overline{c}(j) \!-\!\underline{\alpha_k}^*(j) \!+\!\overline{\alpha_k}^*(j)\!-\!\lambda_k^*\!+\! S_{G,k}(\cdot,m_j)^T\mu_k^*\notag\\
    &=0,
	\end{align}
	\begin{align}
	\label{L/e}
	&\frac{\partial \mathcal{L}_{\uppercase\expandafter{\romannumeral2}}}{\partial \delta g^D_k(j)}\!= \!-\!\epsilon_k\underline{c}(j)\!-\! \underline{\beta_k}^*(j)\!+\! \overline{\beta_k}^*(j) \!+ \!\lambda_k^*\!-\! S_{G,k}(\cdot,m_j)^T\mu_k^*\notag\\
	&\quad \quad \quad \ \ = 0.
	\end{align}
    With $\delta d_k(l),\delta g^U_k(j),\delta g^D_k(j)$ respectively multiplied to both the left-hand side and the right-hand side of (\ref{L/a})-(\ref{L/e}), combined with the complementary slackness of (\ref{dg1fanwei})-(\ref{dk1fanwei}) we have:
	\begin{align}
	\label{L/a1}
	 \lambda_k^* \delta d_k(l) =\epsilon_k c_{L}(l) \delta d_k(l)
	  + S_{D,k}(\cdot,m_l)^T\mu_k^* \delta d_k(l),
	\end{align}
	\begin{align}
	\label{L/b1}
	\overline{\alpha_k}^*(j)r_U(j)\!=\!- \epsilon_k\overline{c}(j)\delta g^U_k(j)\!+\! (\lambda_k^*\!-\!S_{G,k}(\cdot,m_j)^T\mu_k^*)\delta g^U_k(j),
	\end{align}
	\begin{align}
	\label{L/e1}
	\overline{\beta_k}^*(j)r_D(j)= \epsilon_k\underline{c}(j)\delta g^D_k(j)+(S_{G,k}(\cdot,m_j)^T\mu_k^*- \lambda_k^*)\delta g^D_k(j).
	\end{align}
    With (\ref{L/b1}), for generator $j$ we have:
    \begin{align}
	\label{U balancing cost j}
	\Pi^U_k(j)= \overline{\alpha_k}^*(j)r_U(j)+\epsilon_k\overline{c}(j)\delta g^U_k(j)=\omega_k(m_j)\delta g^U_k(j),
	\end{align}
    Similarly, for generator $i$ we have:
    \begin{align}
	\label{U balancing cost i}
	\Pi^U_k(i)=\omega_k(m_i)\delta g^U_k(i).
	\end{align}
    In addition, apparently we know that $\omega_k(m_i)=\omega_k(m_j)$, then with (\ref{U balancing cost j})-(\ref{U balancing cost i}), we can prove Theorem 2.
    
	\section{Proof of Theorem 3 (Individual Rationality)}
	\label{Sec:Individual Rationality}
			\begin{figure*}[!t]
		\normalsize
		\setcounter{MYtempeqncnt}{\value{equation}}
		\begin{align}
		\label{Lagrangian}
		\mathcal{L}(\cdot) =&c^T_{g} g + c_{U}^T r_U + c^T _D r_D + \sum\limits_{k\in \mathcal{K}} \epsilon_k (\overline{c}^T \delta g^U_k -\underline{c}^T\delta g^D_k +c^T_{d}\delta d_k +\lambda(\sum d - \sum g)  + \mu(S_G^Tg-S_D^Td - f)\notag\\& +\overline{\upsilon}(g + r_U - \overline{G}) \!+\! \underline{\upsilon}(\underline{G}+r_D - g) \!+\! \underline{\rho^U}(0 - r_U) \!+\! \overline{\rho^U}( r_U-\overline{r_U}) \!+\! \underline{\rho^D}(0 - r_D)\!+\! \overline{\rho^D}(r_D - \overline{r_D})\! +\!\sum\limits_{k\in \mathcal{K}} \underline{\alpha_k}(0 - \delta g^U_k) \notag\\& + \sum\limits_{k\in \mathcal{K}} \overline{\alpha_k}(\delta g^U_k - r_U ) +\sum\limits_{k\in \mathcal{K}} \lambda_k(\sum (d+\pi_k -\delta d_k) - \sum(g+\delta g^U_k-\delta g^D_k)) + \sum\limits_{k\in \mathcal{K}} \underline{\tau_k}( 0 - \delta d_k) + \sum\limits_{k\in \mathcal{K}} \overline{\tau_k}(\delta d_k - d-\pi_k ) \notag\\ & + \sum\limits_{k\in \mathcal{K}} \overline{\beta_k}(\delta g^D_k - r_D )   + \sum\limits_{k\in \mathcal{K}} \underline{\beta_k}( 0 - \delta g^D_k) +\sum\limits_{k\in \mathcal{K}} \mu_k(S_{G,k}^T((g+\delta g^U_k-\delta g^D_k)-S_{D,k}^T(d+\pi_k -\delta d_k))-f_k).
		\end{align}
		\hrulefill
		\vspace*{4pt}
	\end{figure*}
	We present the Lagrangian of the proposed model (II) in (\ref{Lagrangian}). Furthermore, the Lagrangian of the profit maximization model (IV) of each generator $j$ is
	\begin{align}
	\label{Lagrangian IV}
	\mathcal{L}_{\uppercase\expandafter{\romannumeral4}}&=-\eta^g(j) \times g(j) - \eta^U(j) \times r_U(j) - \eta^D(j) \times r_D(j)
	\notag\\& \quad +c_{g}(j) \times g(j) + c_U(j) \times r_U(j) + c_D(j)\times r_D(j)
	\notag\\& \quad +\overline{\upsilon}(j)(g(j)\!+\!r_U(j)\!-\!\overline{G}(j))\!+\!\underline{\upsilon}(j)(r_D(j)\!-\!g(j))
	\notag\\&\quad +\underline{\rho^U}(j)(0-r_U(j))+\overline{\rho^U}(j)(r_U(j)-\overline{r_U}(j))
	\notag\\&\quad +\underline{\rho^D}(j)(0-r_D(j))+\overline{\rho^D}(j)(r_D(j)-\overline{r_D}(j)).
	\end{align}
	With the formulations of $(\eta^U(j)\!,\!\eta^D(j)\!,\!\eta^D(j))$ in (\ref{RGMP2}), (\ref{RUMP}) and (\ref{RDMP}), we can observe that the Lagrangian of model (IV) in (\ref{Lagrangian IV}) is a part of the Lagrangian of model (II) in (\ref{Lagrangian}). Since these two models are both LP models, according to the KKT conditions we can conclude that, for each generator $j$, its optimal energy and reserve procurement $g^*(j)\!,r_U^*(j)\!,r_D^*(j)$ solved from model (II) is also the solution to its profit maximization model (IV), which proves Theorem 3.
		
	\section{Proof of Theorem 4 (Revenue Adequacy)}
	\label{Sec:Revenue Adequacy}
	To prove revenue adequacy, the phase angle-based form of the proposed co-optimization model is presented as follows:

	\begin{align}
	&\mbox{(V)}: \quad \underset{\{g,r_U,r_D,\delta g^U_k,\delta g^D_k,\delta d_k,\theta,\theta_k\}}{\rm minimize} F^{V}(\cdot), 
	\notag\\ &\mbox{subject to}\notag\\
	&(\ref{base balance and pf}),(\ref{base physical limit1}),(\ref{base physical limit2}),(\ref{dg1fanwei}),(\ref{dg2fanwei}),(\ref{dk1fanwei})\notag\\
	\label{basecase balance2}
	&\Lambda: A_G \cdot g- A_D \cdot d = B \theta,\\
	\label{basecase pf2}
	&\mu: F \theta \leq f,\\
	&\mbox{for all $k\in \mathcal{K}$:}\notag\\
	\label{cntg balance2}
	&\Lambda_k:A_G(g+\delta g^U_k - \delta g^D_k)- A_D(d+\pi_k -\delta d_k)=B_k \theta_k,\\
	\label{cntg pf2}
	&\mu_k:F_k \theta_k \leq f_k,
	\end{align}
	where $F^{V}(\cdot)$ represents the objective function of the phase angle-based model (V), which is the same as $F^{II}(\cdot)$ in shift factor-based model (II). $A_G$ and $A_D$ are matrices that connect generators and loads with nodes, respectively. Vectors $\theta$ and $\theta_k$ are the phase angle vectors in the base case and in scenario $k$, respectively. Matrices $B$ and $B_k$ denote the coefficient matrices of DC power flow equations in the base case and in scenario $k$, respectively. Matrices $F$ and $F_k$ are the branch-node admittance matrices in the base case and in scenario $k$, respectively. With the equivalence of the shift factor-based model (II) and the phase angle-based model (V), for loads we have
    \begin{equation}
	\label{cntgptdf}
	A_D^T\Lambda= \lambda^* \cdot \mathds{1}_{ND \times 1}\!-\!S_D^T\mu^*,A_D^T\Lambda_k = \lambda_k^* \cdot \mathds{1}_{ND  \times 1}\!-\!S_{D,k}^T\mu_k^*,
    \end{equation} 
    and for generators we have
    \begin{equation}
	\label{cntgptdf2}
	A_G^T\Lambda= \lambda^* \cdot \mathds{1}_{NG \times 1}\!-\!S_G^T\mu^*,A_G^T\Lambda_k = \lambda_k^* \cdot \mathds{1}_{NG  \times 1}\!-\!S_{G,k}^T\mu_k^*,
    \end{equation} 
	Where $ND$ and $NG$ are the numbers of loads and generators, respectively. In addition, we denote the Lagrangian of model (V) as $\mathcal{L}_{V}$, with the KKT conditions we have:
	\begin{align}
	\label{baseCR}
	&\theta^T\frac{\partial \mathcal{L}_{\uppercase\expandafter{\romannumeral5}}}{\partial  \theta}=(B\theta)^T\Lambda^* + (F\theta)^T\mu^* = 0,\\
	\label{cntgCR}
	&\theta_k^T\frac{\partial \mathcal{L}_{\uppercase\expandafter{\romannumeral5}}}{\partial  \theta_k}=(B_k\theta_k)^T\Lambda_k^* + (F_k\theta_k)^T\mu_k^* = 0.
	\end{align}
	In addition, for the base case we have:
	\begin{equation} 
	\Gamma^d_0-\Gamma^g_0=(\lambda^* \cdot \mathds{1}_{ND \times 1}\!-\!S_D^T\mu^*)^Td - (\lambda^* \cdot \mathds{1}_{NG \times 1}\!-\!S_G^T\mu^*)^Tg
	\end{equation} 
	and $(f^T\mu=\Delta_0)$. Considering revenue adequacy for the base case, we have:
	\begin{align}
	\label{lemma4}
	&(\lambda^* \cdot \mathds{1}_{ND \times 1}\!-\!S_D^T\mu^*)^Td - (\lambda^* \cdot \mathds{1}_{NG \times 1}\!-\!S_G^T\mu^*)^Tg\notag\\
	=&(A^T_D\Lambda^*)^Td -(A^T_G\Lambda^*)^Tg =\!-\!(\Lambda^*)^T(B\theta)\!=\!(F\theta)^T\mu^*\!=\!f^T\mu^*.
	\end{align}
    These four equations are based on equations (\ref{cntgptdf})-(\ref{cntgptdf2}), the complementary slackness of (\ref{basecase balance2}), equation (\ref{baseCR}), and the complementary slackness of (\ref{basecase pf2}), respectively. With above equations, we have
    \begin{align}
    \label{RA for the base case}
    \Gamma^d_0=\Gamma^g_0+\Delta_0,
    \end{align}
    which proves \textit{revenue adequacy for the base case}.
    
    In addition, the congestion rent contributed from any non-base scenario $k$ $(\Delta_k=f_k^T\mu^*_k)$ is:
	
	\begin{align}
	\label{cntgCR2}
	&\quad f_k^T\mu_k^* = (F_k\theta_k)^T\mu_k^* =-(B_k\theta_k)^T\Lambda_k^*
	\notag\\&=(A_D^T\Lambda_k^*)^T(d+\pi_k -\delta d_k)-(A_G^T\Lambda_k^*)^T(g+\delta g^U_k-\delta g^D_k)
	\notag\\&=(\lambda_k^* \cdot \mathds{1}_{ND  \times 1}\!-\!S_{D,k}^T\mu_k^*)^T(d+\pi_k -\delta d_k)\notag\\
	&\quad-(\lambda_k^* \cdot \mathds{1}_{NG  \times 1}\!-\!S_{G,k}^T\mu_k^*)^T(g+\delta g^U_k-\delta g^D_k)
	\notag\\&=(- S_{D,k}^T\mu_k^*)^T(d+\pi_k-\delta d_k)\!-\!(- S_{G,k}^T\mu_k^*)^T(g\!+\!\delta g^U_k\!-\!\delta g^D_k)
	\notag\\&=(- S_{D,k}^T\mu_k^*)^T(d+\pi_k)+(S_{D,k}^T\mu_k^*)^T\delta d_k +(S_{G,k}^T\mu_k^*)^T g
	\notag\\&\quad+(S_{G,k}^T\mu_k^*)^T\delta g^U_k-(S_{G,k}^T\mu_k^*)^T\delta g^D_k.
	\end{align}
	These six equations are based on the complementary slackness of (\ref{cntg pf2}), equation (\ref{cntgCR}), the complementary slackness of (\ref{cntg balance2}), equations (\ref{cntgptdf})-(\ref{cntgptdf2}), the complementary slackness of (\ref{cntg balance}), and some reorganizations, respectively. On top of that, with the complementary slackness of (\ref{cntg balance}) we have:
	
	\begin{align}
	\label{RA2 middle}
	&\sum_j\lambda_k^*(g(j)+\delta g^U_k(j)-\delta g^D_k(j)))
	\notag\\=&\sum\limits_l \lambda_k^* (d(l)+\pi_k(l) -\delta d_k(l)).
	\end{align} 
	Moreover, with equation (\ref{L/a1}) we have:
	\begin{align}
	&\sum\limits_l \lambda_k^* (d(l)+\pi_k(l) -\delta d_k(l))
	\notag\\=&\sum\limits_l(\lambda_k^* d(l)\!+\!\lambda_k^* \pi_k(l) \!-\!\epsilon_k c_{L}(j) \delta d_k(l)\!-\!S_k(\cdot,m_l)^T\mu_k^* \delta d_k(l)),
	\end{align} 
	which can be reorganized according to (\ref{L/b1})-(\ref{L/e1}) and (\ref{RA2 middle}) as follows:
	\begin{align}
	\label{RA2}
	&\quad \sum\limits_l(\lambda_k^* (d(l)+\pi_k(l))-(\epsilon_kc_L(l)+S_{D,k}(\cdot,m_l)^T\mu_k^*)\delta d_k(l)\notag\\& \quad-\sum\limits_j\lambda_k^* g(j)
	\notag\\&=\sum\limits_j (\overline{\alpha}_k^*(j)r_U(j)+\epsilon_k \overline{c}(j) \delta g^U_k(j)+S_{G,k}(\cdot,m_j)^T \mu_k^* \delta g^U_k(j))
	\notag\\&\quad-\!\sum\limits_j (\!-\overline{\beta}_k^*(j)r_D(j)\!+\!\epsilon_k \underline{c}(j)\delta g^D_k(j)\!+\!S_{G,k}(\cdot,m_j)^T \mu_k^*\delta g^D_k(j)).
	\end{align}
    If we add term $(\sum\limits_l\!S_{D,k}(\cdot,m_l)^T\! \mu_k^*(d(l)\!+\!\pi_k(l)) \!+\! \sum\limits_j\!S_{G,k}(\cdot,m_j)^T\!\mu_k^*g(j))$ and its opposite to the right-hand side of (\ref{RA2}) and reorganize the equation, we have:
	\begin{align}
	\label{RA3}
	&\quad\sum\limits_l(\lambda_k^* \!-\!S_{D,k}(\cdot,m_l)^T\mu^*_k)(d(l)+\pi_k(l))
	\notag\\&=\sum\limits_j(\lambda_k^* g(j)-S_{G,k}(\cdot,m_j)^T \mu_k^* g(j))	
	\notag\\&\quad+\sum\limits_j\overline{\alpha}_k^*(j)r_U(j)+\sum\limits_j \overline{\beta}_k^*(j)r_D(j)
	\notag\\&\quad\!+\!\sum\limits_j \epsilon_k \overline{c}(j)\delta g^U_k(j)\!-\!\sum\limits_j \epsilon_k \underline{c}(j)\delta g^D_k(j)\!+\!\sum\limits_l \epsilon_k c_L(l) \delta d_k(l)
	\notag\\&\quad +(\sum\limits_j S_{G,k}(\cdot,m_j)^T \mu_k^* g(j) + \sum\limits_j S_{G,k}(\cdot,m_j)^T \mu_k^* \delta g^U_k(j)
	\notag\\&\quad - \sum\limits_j S_{G,k}(\cdot,m_j)^T \mu_k^* \delta g^D_k(j) + \sum\limits_l S_{D,k}(\cdot,m_l)^T \mu_k ^*\delta d_k(l)) \notag\\&\quad - \sum\limits_l S_{D,k}(\cdot,m_l)^T\mu_k^*(d(l)+\pi_k(l)).
	\end{align}
    The term on the left-hand side of (\ref{RA3}) is the contribution of scenario $k$ to load payment, including energy payment (\ref{non-base scenario prices to load payment in all scenarios}) and load fluctuation payment (\ref{fluctuation payment}). The right-hand side of equation (\ref{RA3}) include the contribution of scenario $k$ to energy credit (\ref{non-base scenario prices to generator energy credit}) in the $1^{st}$ row, its contribution to upward and downward reserve credit (\ref{upward reserve payments})-(\ref{downward reserve payments}) in the $2^{nd}$ row, its contribution to expected re-dispatch payment (\ref{upward redispatch payments})-(\ref{downward redispatch payments}) and expected load shedding compensation (\ref{load shedding credits}) in the $3^{rd}$ row, and its contribution to congestion rent in the $4^{th}$-$6^{th}$ rows. Therefore, equation (\ref{RA3}) can prove \textit{revenue adequacy for any scenario $k$}. Along with \textit{revenue adequacy for the base case} in (\ref{RA for the base case}), we can prove Theorem 4.

	%
	
	
	

\end{document}